%% file: main.tex
\documentclass[10pt,twocolumn,letterpaper]{article}
\usepackage{usenix-2020-09}

\newif\ifusenixsty
\usenixstytrue
\newif\ifcameraready
\camerareadyfalse
\newif\ifmarkchanges
\markchangesfalse
\newif\ifremovecomments
\removecommentsfalse
\newif\ifshowadditions
\showadditionsfalse
\newif\ifshowrr
\showrrfalse

\input{0_packages}
\input{0_config}
\input{0_macros}

\definecolor{mycitegreen}{RGB}{0,90,0}
\definecolor{myrefblue}{RGB}{90,0,0}

\begin{document}

\pagestyle{empty}


\title{Enhancing Network Failure Mitigation with Performance-Aware Ranking}

\let\textcircled=\pgftextcircled

\ifshowrr
    \input{reviewer_response}
    \clearpage
    \setcounter{section}{0}
    \renewcommand*{\theHsection}{chY.\the\value{section}}
    \setcounter{figure}{0}
\fi


\author{
{\rm Pooria Namyar$^{\ddag}$, Arvin Ghavidel$^{\ddag}$, Daniel Crankshaw$^{\dag}$\thanks{The author contributed to this work while at Microsoft.}~, Daniel S. Berger$^{\dag}$,} \\ 
{\rm Kevin Hsieh$^{\dag}$, Srikanth Kandula$^{\dag}$, Ramesh Govindan$^{\ddag}$, Behnaz Arzani$^{\dag}$}\vspace{1mm}\\
$^{\ddag}$University of Southern California, $^{\dag}$Microsoft
}



\maketitle

\input{abstract}

\input{intro_ramesh2}
\input{case_for}
\input{design_ramesh}

\input{evaluation}
\input{discussion}
\input{relatedWork}

\vspace{1mm}
\parab{Acknowledgments.} \cradd{We thank our shepherd, Sergey Gorinsky, and the anonymous reviewers for their insightful comments. We also thank Ricardo Bianchini, Ranveer Chandra, Jitu Padhye, Solal Pirelli, Rachee Singh and Lihua Yuan for their helpful feedback. This material is based upon work supported in part by the U.S. National Science Foundation under grant No. CNS-1901523.}

\clearpage
\newpage
\bibliographystyle{plain}
\bibliography{reference}

\clearpage
\input{appendix}

\end{document}

%% file: 0_packages.tex
\usepackage{lmodern}
\usepackage{amssymb,amsmath,amsfonts}
\usepackage[T1]{fontenc}
\usepackage[utf8]{inputenc}

\usepackage{upquote}

\usepackage{microtype}
\UseMicrotypeSet[protrusion]{basicmath}

\usepackage{hyperref}
\usepackage{graphicx,grffile}

\usepackage{times} 
\usepackage{tikz}
\usepackage{endnotes,epsfig}
\usepackage{multirow}
\usepackage{xspace}
\usepackage{tabularx}
\usepackage{ragged2e}
\usepackage{booktabs}
\usepackage{paralist}
\usepackage[american]{babel}
\usepackage[shortlabels]{enumitem}
\usepackage{courier,color,wrapfig}
\usepackage{xspace}
\usepackage{balance}
\usepackage{epstopdf}
\usepackage{multirow}
\usepackage{booktabs}
\usepackage{url}
\def\UrlBreaks{\do\/\do-} 
\usepackage{amssymb}
\usepackage{pifont}
\usepackage{subfigure}
\usepackage{tikz}
\usepackage{comment}
\usepackage{mathtools}
\usepackage[normalem]{ulem}
\usepackage{xkeyval}

\ifremovecomments
    \usepackage[disable, textsize=small]{todonotes}
\else
    \usepackage[disable, textsize=small]{todonotes}
\fi

\usepackage{ifthen}
\usepackage{etoolbox}

\usepackage[margin=5pt,font=small,labelfont={bf,rm}]{caption}
\DeclareCaptionFormat{myformat}{#1#2#3\hrulefill}
\usepackage{sepfootnotes}
\usepackage[compact]{titlesec}
\usepackage{bbding}

\usepackage{array}
\newcolumntype{L}[1]{>{\raggedright\let\newline\\\arraybackslash\hspace{0pt}}m{#1}}
\newcolumntype{C}[1]{>{\centering\let\newline\\\arraybackslash\hspace{0pt}}m{#1}}
\newcolumntype{R}[1]{>{\raggedleft\let\newline\\\arraybackslash\hspace{0pt}}m{#1}}
\usepackage{amsmath,amsthm}
\usepackage{bbm}
\usepackage[linesnumbered,boxed]{algorithm2e}
\usepackage{setspace}

\makeatletter
\def\BState{\State\hskip-\ALG@thistlm}
\makeatother


%% file: 0_config.tex
\makeatletter
\def\maxwidth{\ifdim\Gin@nat@width>\linewidth\linewidth\else\Gin@nat@width\fi}
\def\maxheight{\ifdim\Gin@nat@height>\textheight\textheight\else\Gin@nat@height\fi}
\makeatother
\setkeys{Gin}{width=\maxwidth,height=\maxheight,keepaspectratio}
\makeatletter
\g@addto@macro{\UrlBreaks}{\UrlOrds}
\makeatother

\everypar{\looseness=-1}

\newcommand\paraspace{\vspace*{.5ex}}
\providecommand\parab[1]{\paraspace\noindent\textbf{#1}}
\providecommand\parae[1]{\paraspace\textbf{\textit{#1}}}

\apptocmd\normalsize{%
\abovedisplayskip=5pt
\abovedisplayshortskip=5pt
\belowdisplayskip=5pt
\belowdisplayshortskip=5pt
}{}{}



%% file: 0_macros.tex
\newcommand{\sysname}{\textsc{swarm}\xspace}
\newcommand{\impact}{CLPEstimator\xspace}
\newcommand{\azure}{Azure\xspace}

\newcommand{\traffic}{\set{T}}
\newcommand{\actionTraffic}{\set{T}_{a}}
\newcommand{\shortflowtraffic}{\set{T}_{s}}
\newcommand{\longflowtraffic}{\set{T}_{l}}
\newcommand{\topology}{G}
\newcommand{\actionTopology}{G_{a}}
\newcommand{\mitigationAction}{\set{M}}
\newcommand{\switchset}{\set{V}}
\newcommand{\linkset}{\set{E}}

\newcommand{\confidenceLevel}{\alpha}
\newcommand{\impactFlows}{\beta}
\newcommand{\epochsize}{\zeta}

\newcommand{\listcurrentflows}{\set{K}}

\newcommand{\throughput}{\theta}

\newcommand{\ie}{\emph{i.e.,}\xspace}
\newcommand{\eg}{\emph{e.g.,}\xspace}

\newcommand{\secref}[1]{\S\ref{#1}}
\newcommand{\figref}[1]{Fig.~\ref{#1}}

\newcommand{\tabref}[1]{Table~\ref{#1}}
\newcommand{\algoref}[1]{Alg.~\ref{#1}}

\newcommand{\lineref}[1]{Line~\ref{#1}}

\newcommand{\behnaz}[1]{\todo[author=BA,color=red!30,inline]{#1}}

\newcommand*\circled[1]{\tikz[baseline=(char.base)]{
            \node[shape=circle,draw,fill=yellow,inner sep=1pt] (char) {#1};}}

\ifmarkchanges
	\newcommand{\cradd}[1]{\textcolor{blue}{#1}}
	\newcommand{\crdel}[1]{\sout{\textcolor{red}{#1}}}
\else
	\ifshowadditions
		\newcommand{\cradd}[1]{\textcolor{blue}{#1}}
	\else
		\newcommand{\cradd}[1]{#1}
	\fi
	\newcommand{\crdel}[1]{}
\fi

\newcommand{\cadd}[1]{#1}

\newcommand{\set}[1]{\mathcal{#1}}

\newcommand{\cut}[1]{}



%% file: reviewer_response.tex
\section*{Review Response}

\newcommand{\quotereview}[1]{\medskip\begin{center}\fbox{\parbox{0.40\textwidth}{\it {#1}}}\end{center}\medskip}

\newenvironment{Itemize}%
{\begin{itemize}
\setlength{\leftmargin}{1em}%
\setlength{\itemsep}{0in}%
\setlength{\topsep}{-.1in}%
\setlength{\partopsep}{-.1in}%
\setlength{\parsep}{-.1in}%
\setlength{\parskip}{-0in}}%
{\end{itemize}}

We thank the reviewers for their time and insightful feedback. We summarize our key changes below and point to specific changes in the submission when responding to the meta-review, the common reviews, and the individual reviews~(\secref{rr:meta},~\secref{rr:common}, and~\secref{rr:individual} respectively). 



\subsection{Summary of Key Changes}
\label{rr:change_list}

Here is the list of major changes:
\begin{Itemize}
    \item We have extended our evaluation to NS3 simulations and to new flow size distributions. See~\secref{rr:ns3}.
    \item We have improved the writing of the evaluation section to summarize the key information regarding our experiment setup. We also add a section to explain all the remaining details in the appendix. See~\secref{sss:mr_c1}.
    \item We added a discussion regarding the generality of \sysname and explain settings where it applies or does not apply. See~\secref{sss:mr_c2}. 
    \item We clarified how today operators gather the necessary inputs for \sysname and how robust \sysname is to these inputs. See~\secref{sss:mr_c3}.
    \item We have rewritten a portion of the paper to explain the intuitions and the reasons behind our model. See~\secref{sss:mr_c4} and~\secref{rr:common2}.
    \item We have also added a discussion on how operators can choose the list of mitigations and a comparator to use \sysname. See~\secref{sss:mr_c5}.
\end{Itemize}

\subsection{NS3 Simulation}
\label{rr:ns3}
We have successfully extended our experiments to NS3 simulations. These allowed us to evaluate \sysname at larger scales than our emulation and testbed experiments. By itself, NS3 takes over a day to complete one simulation run on a 128-server topology. We invested significant effort in parallelizing it using MPI to reduce the time to run one sample to 6 hours. In the end, we show results from experiments that took more than 2100 VM hours. Our results confirm that \sysname finds effective mitigations at larger scales where existing auto mitigation systems can lead to 30\%~--~78\% higher 99p FCTs. We also extend and validate our results to another flow size distribution (see~\secref{s:eval-other}).

\subsection{Meta Review}
\label{rr:meta}

\subsubsection{Comment 1}
\label{sss:mr_c1}
\quotereview{Clarifying Experiment Setup}


We summarize the key parts of our experiment set up in~\secref{sec:evaluation} and extended~\secref{appendix:experiment-details} to explain the rest of the details. Here are the key points:

\parab{Traffic Trace (\secref{a:traffic-trace}).} These refer to a sequence of flows (source, destination, size, and start time) that we want to use to evaluate. We follow common practices~\cite{hpcc,BeyondFattrees,ExpandTime} to ensure these are realistic. Specifically, we (1) ensure trace duration is long enough to represent various flow sizes, (2) ensure we do not capture the impact of empty network, (3) use common and well-known distributions for different aspects such as flow size~\cite{Alizadeh-DCTCP} or flow arrival~\cite{BeyondFattrees,hpcc,ExpandTime,teh2020couder}, and (4) run for many different traffic traces to ensure our evaluation is robust.





\parab{Failure Types (\secref{a:failure}).} We report results on three types of incidents and 57 specific scenarios to evaluate the generality and robustness of \sysname. These incident types are common in \azure and other cloud providers as reported in different case studies~\cite{Arzani-007,omnimon,Roy-NSDI17,Gill-UnderstandingFailures,Zhuo_CorrOpt,Wu_NetPilot}. We leverage the symmetry in datacenters~\cite{Alipourfard_Janus} and choose each of these 57 scenarios to be representative of a wide range of potential incidents (see~\secref{a:failure} for details). In the end, we report over 5 months' worth of Mininet experiments (over 4000 hours) and larger-scale testbed/simulation validations.




\parab{Routing.} We assume the routing is either ECMP or WCMP (common in data centers).

\parab{Emulation setup (Mininet~\secref{a:setup}).} We choose Mininet as our primary evaluation framework because it leverages a real TCP/IP stack from the Linux kernel, enabling evaluation with the production-grade implementation of congestion control algorithms (\eg BBR~\cite{bbr}, Cubic~\cite{cubic}). We explain how we overcome the challenges in using Mininet and the exact set up in~\secref{a:setup}.

\parab{Simulation setup (NS3~\secref{a:setup}).} We added results based on NS3, which helps us simulate larger topologies with higher link bandwidth and lower delays (compared to Mininet experiments). Our results confirm the effectiveness of \sysname.

\parab{Testbed setup (\secref{a:setup}).} We also evaluate \sysname in a testbed to show its effectiveness when dealing with real links, switches, and servers.

\subsubsection{Comment 2}
\label{sss:mr_c2}
\quotereview{Discuss SWARM's generality.}



\parab{Topologies (\secref{sec::discussion}).}
\sysname applies to any datacenter topology that relies on ECMP or WCMP for routing. These are commonly used in Clos and its variants. In fact, we use different variants of Clos in our testbed (the tier-1 switches are connected to all the tier-2 switches), and our simulation/emulation (tier-1 switches in the same pod are connected to different tier-2 switches). In both cases, \sysname finds effective mitigations. However, \sysname does not apply to topologies that use traffic engineering to route the demands (for example, direct-connect-based data center topologies~\cite{teh2020couder}), and we refer to future work to address these. We clarified this in the discussion section (\secref{sec::discussion}).

\parab{Failure and Mitigations (\secref{s:other-details}).} \sysname applies to any incident and any mitigation that we can describe as a change to our network state and traffic representation (this includes all the failures in~\tabref{Table::failurelist}).

\parab{Traffic distributions (\secref{s:eval-other}).} \sysname applies to any traffic distributions (e.g., flow size and flow arrival distributions). It takes these as inputs (which are already collected in cloud providers) and generates random traffic traces from these distributions.

We show \sysname's effectiveness for both DCTCP~\cite{Alizadeh-DCTCP} and FbHadoop~\cite{hadoop-workload}'s flow size distributions in our simulation (These two workloads are different. FbHadoop consists of many short flows whereas DCTCP has more balance between short and long flows). We also show that \sysname is robust to errors in flow arrival rates.


\subsubsection{Comment 3}
\label{sss:mr_c3}
\quotereview{Detection and Troubleshooting of Failures. Sensitivity of \sysname.}



\parab{Detecting \& Troubleshooting Failure.} We discuss this in~\secref{s:sysname-overview} when explaining the inputs to \sysname. In large cloud providers, there are monitoring systems in place to identify and report these incidents~\cite{GCP-Incidents,Azure-RCA}. All the failures that we discussed in \tabref{Table::failurelist} can be detected and reported using existing monitoring systems. Once an incident is detected, either an auto mitigation system or an on-call engineer uses the information in the report to find the proper mitigation to reduce the impact of failures while others work on a repair. \sysname is an auto mitigation system and only applies to the detected incidents. For example, \cite{Zhuo_CorrOpt} reports Microsoft can detect and apply mitigations when the packet drop rate is above a certain threshold ($10^{-6}$). The existing incident reports contain the information \sysname needs (\eg packet drop rate). 

\parab{Sensitivity.} In \secref{sec::sensitivity}, we show that the incident information, such as packet drop rate, does not need to be exact for \sysname to provide effective mitigation.

\subsubsection{Comment 4}
\label{sss:mr_c4}
\quotereview{Consider finer-grained traffic flow classification}

We added a discussion in~\secref{s:approach-challenges}. At a high level, our goal is to abstract routing and transport behavior enough to find effective mitigation quickly for large-scale data centers rather than accurately estimating the flow performance. Existing methods to estimate performance either are too slow~\cite{henderson2008network}, require a prohibitive amount of compute~\cite{mimicnet}, or do not handle failures~\cite{parsimon}. Our binary classification into short and long flows matches general transport protocol behavior (see below) and allows us to be simple enough to scale. Our results already show \sysname is significantly better than other mitigation strategies~\cite {Wu_NetPilot,Zhuo_CorrOpt} that completely ignore the traffic, routing, and failure characteristics. There are several reasons why this classification helps:

\parab{Network Mitigation} impacts long and short flows differently. Short flows only experience a snapshot of the network whereas as large flows stay longer and are affected by the network state variations.

\parab{Transport protocols} treat short flows and long flows differently. They~\cite{bbr,cubic} usually have a start-up phase until they find the bandwidth limit. Short flows with few packets are not long enough to reach their steady state. Prior work has shown that these short flows are impacted by queueing delay rather than bandwidth limit~\cite{Alizadeh-DCTCP,Homa-Montazeri}. Accounting for short flows differently improves the effectiveness of our approach. (see~\secref{rr:common2})

\parab{Datacenter Workloads} are a mixture of delay-sensitive short flows and throughput-sensitive long flows~\cite{Alizadeh-DCTCP,Homa-Montazeri}. \sysname reports performance for each type separately, allowing operators to tailor mitigations based on their requirements and workloads.


\parab{Scalability.} The scalability of \sysname depends on how fast we can compute the fair share of bandwidth for each flow. When excluding short flows, the number of flows we consider decreases, improving the scalability.

We added finer-grained classification as one of the future steps in~\secref{sec::discussion}.

\subsubsection{Comment 5}
\label{sss:mr_c5}
\quotereview{How operators could make use of SWARM}


We have not yet deployed \sysname. However, as we describe in several places in the paper, we have taken care to consult with operators at \azure, and to use information that providers already collect. Furthermore, at \azure, auto-mitigation systems already rank possible mitigations. As such, we expect \sysname to be a relatively easy drop-in replacement for the mitigation-ranking component of an auto-mitigation system, especially since \sysname relies on inputs that are already collected in cloud providers (including \azure).

\parab{List of Mitigation for an Incident.} We discuss this in~\secref{s:sysname-overview}. Cloud providers~\cite{autotsg,tsg} document possible mitigation and troubleshooting guidelines for each type of incident, which are used by on-call engineers (same holds for \azure). We only need users to encode the set of actions from their troubleshooting guidelines as inputs to \sysname.

\parab{Choosing a Comparator.} We added a discussion in~\secref{s:sysname-overview}. Cloud providers can choose the comparator based on the dominant workload in a cluster. If delay-sensitive short flows constitute most of the traffic, they can choose a comparator that prioritizes reducing impact on short flows. \sysname takes the comparator as input and chooses the best mitigation that matches the operator's objective. This flexibility helps cloud providers since they might observe different workloads in different clusters and want to adapt the mitigation accordingly. In contrast, existing auto mitigation systems~\cite{Wu_NetPilot,Zhuo_CorrOpt} choose the same action regardless of the preference. Note that \sysname supports any function on the distribution of flow completion time or throughput as a comparator.  

\subsubsection{Comment 6}
\label{sss:mr_c6}
\quotereview{Consider condensing earlier sections for richer evaluation in main body.}
We changed the writing and reduced the earlier sections to accommodate more evaluation and discussion in the main body.



\subsection{Common Reviews}
\label{rr:common}

\subsubsection{Reviewer A \& B}
\label{rr:common-1}
\quotereview{Negative performance penalty}




We added a discussion on this in the evaluation section (\secref{ss:baseline-comparison}). Often, there is a trade-off where one metric's improvement would come at the cost of another. For instance, in~\figref{fig:scenario_1_results:priorityfct}, the first priority is to reduce the impact on short flows' 99p FCT. As a result, optimal mitigation is an action with lowest impact on FCT even if this means reduced performance on other metrics. As you can see, there are scenarios where CorrOpt-75 selects actions that worsen the FCT metric (highest priority) by up to 80\% compared to the optimal action but with better average throughput (lower priority) than the optimal action.

\subsubsection{Reviewer B \& C}
\label{rr:common2}
\quotereview{Impact of Congestion Control Algorithm}


We expanded the discussion in~\secref{s:approach-challenges} to explain these.
The congestion control algorithm (CCA) itself is not directly an input to \sysname. Rather, the input is a simple abstraction that captures the behavior of the transport protocols well enough to find effective mitigations while scaling to large data centers. This abstraction consists of three parts: (1) assume long flows follow max-min fairness, which matches the objective of TCP~\cite{Ben-StatisticalBWSharing}, (2) long flows are either loss limited or capacity limited (see~\figref{fig:ablation-study}), and (3) short flows do not reach steady-state and are delayed due to packet drops or queueing delays rather than bandwidth limits.

To parametrize this abstraction for different transport protocols, we perform offline measurements~\secref{appendix:offline-measurement} to quantify the impact of congestion control mix on the parameters of our abstraction (throughput of long flow in case they are loss limited and impact of packet drops/queueing delay on short flows). Note that we measure enough samples to ensure our testing is statistically significant.

While this abstraction does not capture the exact interaction between flows, it is enough to find effective mitigation while also scaling to large topologies, as demonstrated in our experiments. We also analyze the sensitivity of \sysname and its mitigations to the choice of congestion control algorithms in~\figref{fig:sensitivity:CC} using BBR and CUBIC that have two completely different behavior under loss.


\subsubsection{Reviewer B \& F}
\label{rr:common3}

\quotereview{Metrics other than Throughput and FCT}




CLP is arguably better than the simplistic metrics used in existing
auto mitigation systems~\cite{Wu_NetPilot,Zhuo_CorrOpt}, such as the number of uplinks or the number of paths. These ignore the characteristics of failures (\eg packet drop rate), the traffic, and the impact on the end-to-end performance (see \tabref{tab:intro-compare-methods}). Our results show that using these leads to sub-optimal mitigations that impact connection-level performance more than they need to.

Cloud providers and operators have often used flow completion time (FCT) and Throughput to measure the performance in datacenters (\eg to evaluate new datacenter topology designs~\cite{ExpandTime,teh2020couder} or congestion algorithms~\cite{Homa-Montazeri,Alizadeh-DCTCP,pfabric}). These evaluate, respectively, the performance properties of importance for short and long flows. Moreover, in today's clouds, tail performance metrics are used to characterize performance objectives, so we chose 1p throughput and 99p FCT as metrics.

As and when clouds expand their performance objectives to other metrics, such as jitter, future work can explore how to incorporate these in \sysname (as we discuss~\secref{sec::discussion}).


\subsubsection{Reviewer C \& E}
\label{rr:common4}

\quotereview{Measuring Throughput under different network conditions.}

We discuss this in~\secref{appendix:offline-measurement}. Our model assumes long flows are either loss-limited or capacity-limited (whichever is smaller, see~\secref{s:approach-challenges}, also verified in~\secref{sec::appendix_ablation}). For the capacity-limited part, we assume TCP-friendliness and compute the rate using a fast max-min fair allocator.

The goal of this offline measurement is to quantify what happens when the flows are loss limited. In these cases, the interaction between the flows and the bandwidth limit matters less. Instead, we need to quantify the rate the control loop converges under different packet drops. To do this, it suffices to use a simple topology (as in~\figref{fig:offline:topo1}) with large link capacities to ensure bandwidth is not a bottleneck. We only need to send one long flow and measure its average throughput under different network setups (packet drop, RTT, and congestion control). For congestion control, we sample from the congestion control mix we observe in the cloud (analyzed the sensitivity in~\ref{a:sensitivity-congestion}). Since this is a small topology, we can create it using the same devices as the target environment (\eg we used Mininet to perform this offline measurement for our Mininet experiments, and use the testbed for our testbed experiments).

We want to emphasize again that we do not want to estimate the performance accurately (over-indexing on specific protocol, while accurate, is slow~\cite{henderson2008network}). Instead, we want to have a model that is accurate enough to rank mitigations while being simple enough to scale to large data centers.


\subsection{Other Individual Reviews}
\label{rr:individual}

\subsubsection{Reviewer A}

\quotereview{Accuracy of \sysname's estimations}

We want to emphasize that the goal of \sysname is \textit{not} to estimate the performance of flows accurately but to ensure we can 
find high-quality mitigation actions quickly as scale. Given that in practice there are likely to be a few mitigation choices with widely varying impact, a mitigation ranking system can tolerate errors in performance estimates while still being able to rank the alternatives effectively. We exploit this property to design approximate but scalable performance estimators.

Regardless, in~\figref{fig:ablation:multi-epoch-multi-sample}, we show the impact of our design choices on the relative error of throughput estimates with respect to our Mininet measurements. 


\quotereview{Clarifying Scalability Experiment in~\figref{fig:scalability}}


We changed the writing in~\secref{s:eval-other} and caption of~\figref{fig:scalability} to clarify the scalability experiments.

\parab{Error discussed in \figref{fig:scalability:error}.} In~\secref{s:other-details}, we explain our different techniques to scale \sysname. Among those, three of them can introduce additional errors (\ie an approximate max-min fair algorithm, reducing the number of epochs, and traffic down scaling). In~\figref{fig:scalability:error}, our goal is to quantify the error introduced by these techniques. Therefore, we compare them with respect to a version of \sysname that uses an exact max-min fair algorithm~\cite{Lavanya-sPerc} and does not perform any of these scaling techniques. We observe that these methods substantially speed up the process while having little to no impact on the final estimation. For example, the approximate max-min fair algorithm introduces $<0.9\%$ error compared to using an exact algorithm. 



\parab{Results for 16$K$ servers topology.} In~\figref{fig:scalability:run-time}, our goal is to show how long it takes for \sysname to finish the computation and come up with a mitigation action at a large scale. Our simulations and emulations go a step further; they apply the mitigation to the simulated or emulated network to measure the actual impact on CLP metrics. For this reason, the simulations and emulations do not scale to large topologies, but, as~\figref{fig:scalability:run-time} shows, \sysname's mitigation ranking does.


\quotereview{Improvement of \sysname}

In general, the cause of \sysname's benefit depends on the scenario. In some incidents, \sysname chooses the actions that are already supported but ignored by prior work. In some other cases, the benefit is due to the larger action space of \sysname. We added a discussion with examples in~\secref{a:benefit-two-ex}. 

We emphasize that \sysname's larger action space is a consequence of being able to explore CLP metrics. For example, existing methods~\cite{Wu_NetPilot,Zhuo_CorrOpt} can not reason about the impact of packet drop rates and the trade-off between congestion-induced versus failure-induced packet drops. Therefore, they cannot evaluate the performance of new routing (WCMP) in the presence of packet drops.


\quotereview{Number of Mininet experiments}

We exploit the symmetry in datacenter topologies~\cite{Alipourfard_Janus} and choose these 57 scenarios to be representative of a much broader range of incidents. For example, we can capture any single-link failure using two scenarios. Please refer to~\secref{a:failure} for an extended discussion on this point.

\quotereview{Whether Downscaling is Realistic}


Indeed, to address the point raised, we validated our results in a physical testbed consisting of links with 10$Gbps$ bandwidth and $200\mu s$ propagation delay. We have also now added NS3 simulations with a link bandwidth of 20Gbps and a link delay of 100$\mu s$. We still observe that \sysname can find effective mitigations.

\quotereview{Simulations instead of emulations}

Thanks for bringing this up. We extended our evaluation to NS3 simulations to show how \sysname performs at a larger scale.

\quotereview{Mininet Topology}

We mentioned this in the evaluation section but improved the writing to clarify. Please refer to~\secref{sss:mr_c1} for more information.

\quotereview{Negative Performance Penalty}

Please refer to~\secref{rr:common-1}.

\subsection{Reviewer B}
\quotereview{Other metrics}

Please refer to~\secref{rr:common3}

\quotereview{Biasing toward specific flow sizes}

We do not bias the mitigation toward flows with specific sizes. We provide the users of our system the option of choosing the desired comparator based on their workloads. Then, we decide which mitigation is effective based on that comparator. For example, if operators choose to prioritize short flow's FCT compared to large flow's throughput or vice-versa, \sysname is flexible enough to adjust the mitigation accordingly (see~\figref{fig:scenario_1_actions}). In contrast, existing mitigation systems choose the same action regardless of the preference~\cite{Wu_NetPilot,Zhuo_CorrOpt}. We also support the weighted combination as a comparator (\secref{appendix:other-comparators}).

\quotereview{Traffic demands across 5 minutes}

Please refer to~\secref{rr:meta} for the description of how we generate traffic traces in our evaluations. We ensure the duration of the trace is long enough to capture enough samples from various flow sizes (5 5-minute duration was enough in our Mininet experiments).

\quotereview{Logic behind disabling a link in response to congestion above ToR}

We added a discussion in~\secref{a:logic-mitigation} and a summary in the caption of \tabref{Table::failurelist}. This is a fairly subtle point, thanks for catching this. One of the reasons behind congestion is routing protocols used in datacenters (ECMP/WCMP) often ignore the asymmetry in the network. A common way to deal with this is to disable the congested link so that the routing can utilize other paths. For example, each logical link between two switches (also known as LAG~\cite{Wu_NetPilot}) consists of multiple physical links. LACP is the control protocol on top of LAG that multiplexes packets over these physical links by hashing their header. Each of these physical links can face cuts, resulting in reduced link capacity. If the load is high enough and the link capacity becomes low enough, we will have congestion on that logical link. To deal with this, we can disable the logical link so ECMP/WCMP can utilize other paths with healthy links to route the packets.

\quotereview{Action space of CorrOpt and NetPilot}

We cover all the failures and mitigations in CorrOpt~\cite{Zhuo_CorrOpt} and in NetPilot~\cite{Wu_NetPilot}. Broadly, \sysname applies to any failure and mitigation that we can model as changes to our network state and traffic representation (see \tabref{Table::failurelist} and \secref{s:other-details}).

\quotereview{Parameter space of \sysname.}

\parab{Sensitivity to Inputs.} \sysname takes parameters of the traffic distribution (\eg flow arrival rate) and the characteristics of the failure as inputs (\eg drop rate). In~\secref{sec::sensitivity}, we analyze the sensitivity of \sysname to these inputs. We find \sysname is robust and works well even if there are errors in these measured inputs. For example, we find the error in packet loss rate has to be an order of magnitude for \sysname to make the wrong decision, an unlikely possibility in today's networks. 

\parab{\sysname's Parameters.} \sysname also has some parameters:

\parae{Number of samples.} The number of samples and simulations should be chosen in a way that provides statistical significance (we chose it based on DKW inequality~\cite{DKWInequality} as we explain in~\secref{sec:swarm:internals} under "routing uncertainty" and "traffic variability").

\parae{Epoch Duration.} The throughput of the long flows changes every time a new flow arrives or departs. Therefore, an ideal epoch duration should be such that only one of those events happens in every epoch. For example, the flow inter-arrival rate in our Mininet experiments (after downscaling) is 1 ms (1000 flows arrive per second), and therefore, the epoch duration ideally should be less than 1ms. However, our goal is not to estimate the performance of flows accurately but rather to find effective mitigations. Therefore, we can tolerate errors. In our Mininet experiments, we use 200ms as our epoch size (200$\times$ larger than the inter-arrival time of flows) and still find close-to-optimal mitigations. We also swept the epoch size values from 120ms to 300ms, and we did not observe any change in the suggested mitigation. 

\parae{Number of epochs.} The number of epochs is determined by the trace duration (trace duration = epoch duration $\times$ number of epochs). We should set the trace duration long enough to capture the performance of various flow sizes. Since traces are generated offline, we can measure the number of flows and tune the trace duration accordingly. 


\quotereview{Congestion control as input to \sysname}

Please refer to~\secref{rr:common}.

\quotereview{Routing Samples}

We compute the probability of a flow taking each path using WCMP weights (\figref{fig:routing-prob}) and generate random samples based on this distribution.  This ensures our choice of mitigation is robust to routing samples that constitute the tail performance.

\quotereview{$50\mu s$ propagation delay in Mininet}

Emulating 50$\mu s$ links is impossible in Mininet. That is why we downscale the traffic and the capacities in a way that we can faithfully measure the performance while meeting Mininet's limits (see~\secref{sec:eval-meth} and~\secref{a:setup}). We also have testbed and NS3 experiments that capture cases where links have micro-second level propagation delay ($200\mu s$ and $100\mu s$, respectively). 

\quotereview{Negative Penalties}
Please refer to~\secref{rr:common}. We also updated the text and clarified it in the main body of~\secref{ss:baseline-comparison}.

\quotereview{Clarification about testbed experiments}

\parab{Packet Drops.} Since we are artificially injecting packet drops using ACL, the packet drops are limited by the hardware and have to be the power of 2 (see~\secref{sec:eval-meth} and~\secref{a:exp-other-details}). We consider one high packet drop rate and one the lowest possible packet drop rates we can inject.

\parab{Failure Scenario.} Our goal in our testbed experiment is to show that \sysname can find effective mitigations when dealing with real devices (links, switches, and servers). Therefore, we choose a scenario where two links at two different levels of the data center start dropping packets at two different rates. In this example, we can see the trade-off between causing packet drops by taking no action and causing network congestion by disabling the links (We clarify this in~\secref{s:eval-other}). \sysname was able to find effective mitigations, with $\leq 0.9\%$ performance penalty compared to the optimal mitigation.

\parab{Mitigation Actions.} In this experiment, we consider four mitigation strategies due to hardware limitations (disable each faulty link or take no action. We clarify this in~\figref{fig:testbed_scenario}). \figref{fig:testbed_scenario} shows the performance penalty for \sysname compared to the optimal strategy as well as the worst action (to see how well \sysname is performing). Note that the optimal mitigation has zero performance penalty. The fourth mitigation has a similar performance to the worst case since they choose to keep the faulty link with a high drop rate. Note that the operator solution as well as~\cite{Wu_NetPilot,Zhuo_CorrOpt} would also decide to keep the high-drop link in place to avoid congestion, which would result in these worst-case behaviors. These existing methods have static thresholds that would ignore the characteristics of the failure (\eg packet drop rate) and the network load when making decisions.


\subsubsection{Reviewer C}

\quotereview{Clarifying the example in~\figref{fig:example_two_drops_motiv}}

\parab{Fiber cut.} In~\figref{fig:example_two_drops_motiv}, the fiber cut causes the link to stop forwarding any packets. This reduces the capacity of the network, and the impacted flows should utilize alternative links that already have some traffic. The increased load on the remaining links would cause congestion. NetPilot and Operators disable the congested link, which would reduce the network capacity even more (\secref{a:logic-mitigation})


\parab{NetPilot scenario.} We changed the writing to clarify this. NetPilot would disable the faulty link in both high FCS and low FCS scenarios. While this is an effective mitigation when the network is under low utilization or the packet drop rate is high (High FCS), it significantly impacts the performance in case of low FCS (when the impact of congestion is higher than the low FCS). Our goal with this example is to show that a fixed mitigation strategy that ignores the network load and the failure characteristics (\eg packet drop rate) is not general and can negatively impact the performance.

\quotereview{Numbers reported in~\secref{sec:motivation} and details about \azure}

We clarify this in~\secref{sec:motivation}. The numbers reported are based on our Mininet experiments. As explained in~\secref{sec:eval-meth} and~\secref{a:exp-other-details}, we had to downscale the capacity and the traffic in Mininet, which is why the numbers are low. Regarding details of \azure, we are happy to provide them to PC chairs to avoid violating the anonymization policy.

\quotereview{Clarifying \figref{fig:example_num_active_long_flows}}

We clarify the topology and the incident in the caption of \figref{fig:example_num_active_long_flows}. The topology is similar to~\figref{fig:example_two_drops_motiv}, but the failure scenario only consists of a single T0-T1 dropping packets. The low and high drop rates refer to Low and High FCS in the motivation section ($O(10^{-5})$ and $O(10^{-2})$ drop rate. When we disable a link, the capacity of the network drops and more flows have to share links, which results in these flows receiving a lower rate and staying longer in the network. Therefore, these flows over time accumulate and increase the number of active flows in the network.

\quotereview{Measuring Average Throughput under different conditions.}
Please refer to~\secref{rr:common4}


\quotereview{Impact of congestion control}

Please refer to \secref{rr:common}.

\quotereview{Increase failure tolerance threshold to take no action}

Note that low packet drop rates can impact delay-sensitive short flows, and therefore, it is important to capture and mitigate these. For example, Microsoft~\cite{Zhuo_CorrOpt} detects any links with a drop rate above $O(10^{-6})$ and disables it after safety checks. Our goal is to ensure that we choose mitigations that effectively reduce the impact of the incident. In cases where the network load is low, this involves disabling a low drop rate link. On the other hand, if the network load is high, the impact of congestion can be higher than the drop rate, and therefore, taking no action might be a better choice. \sysname is able to consider all these factors and suggest effective mitigations, as we demonstrated in our experiments.

\quotereview{Downscaling process in Mininet.}

We decrease the link capacities, increase the link delays, and increase the inter-arrival times by the same factor~\cite{Konstantinos_downscale_simulation,Shrink-Pan}. Please see~\secref{rr:meta}.

\subsubsection{Reviewer D}


\quotereview{Classification of flows}

Please refer to~\secref{sss:mr_c4}.

\quotereview{Choosing the comparator.}

Please refer to \secref{sss:mr_c5}.

\subsubsection{Reviewer E}

\quotereview{Generality of SWARM and extending to other cases}

Please refer to~\secref{rr:meta} for a discussion on generality. In summary, \sysname applies to any topology that uses ECMP/WCMP. In fact, we use the variant of Clos in our testbed, in which \sysname finds effective mitigations.

\quotereview{Failures with small impact}

We want to answer a broader question of what happens if the distribution of two mitigations is very similar. In that case, choosing either of the actions would lead to a similar performance. Therefore, choosing the optimal mitigation becomes less important compared to cases where the difference is significant. For example, \figref{fig:scenario_1_results:priorityfct} shows cases where \sysname was not able to clearly distinguish between two choices of mitigations but still has $\leq 5\%$ performance penalty, much better than prior work. 

\quotereview{Dynamic Routing and link capacity}

We currently support fixed routing and link capacities (which are derived from the physical topology, the incident, and the mitigation). We leave this extension to future work. Since \sysname has an epoch-based approach, this involves encoding the changes in routing or link capacities from one epoch to the next such that the approach remains scalable.

\quotereview{Mitigation and Failure List}

\parab{Candidate Mitigations.} Please refer to~\secref{sss:mr_c5}.


\parab{Constructing \tabref{Table::failurelist}.} Please refer to \secref{rr:meta}. We based \tabref{Table::failurelist} on common failures in \azure and other cloud providers~\cite{Arzani-007,Roy-NSDI17,Gill-UnderstandingFailures,Zhuo_CorrOpt,Wu_NetPilot}. \sysname is not limited to this list and supports any failure and mitigation if we can describe it in terms of a change in our network state or traffic representation (see~\secref{sec:swarm:internals} and~\secref{s:other-details}).

\quotereview{Measure throughput under different network conditions}
Please refer to~\secref{rr:common4}.

\quotereview{Clarifying Improvement of \sysname}

\figref{fig:enter-label} shows the distribution of the performance penalty of \sysname and other baselines in terms of their 99p short flow FCT (this corresponds to~\figref{fig:scenario_1_results:priorityfct}). The average performance penalty of \sysname is 0.03\%, whereas the smallest average penalty among all the other baselines is 35.06\%. The maximum performance penalty of \sysname is 0.1\%, whereas the smallest maximum penalty among all the baselines is 79.3\% (hence, 793$\times$).

\quotereview{Implementing in C instead of python}

Thanks for this suggestion. We will consider this in the future. Please note that our code is currently a research prototype that scales well~\ref{fig:scalability:run-time}. 

\quotereview{Clarifying \sysname Experiment Set up}

Please refer to \secref{sss:mr_c1}

\quotereview{Realistic Failure Scenarios}

Please refer to~\secref{rr:meta}.

\quotereview{Discuss SWARM robustness}

Our results in~\secref{sec:evaluation} show that \sysname is able to find better mitigations across three common types of failures under different conditions (\eg drop rate). This indicates \sysname is more robust compared to previous threshold-based methods that choose mitigation based on a static, predetermined threshold, which is not effective in certain conditions (see~\secref{sec:motivation}). 

\subsubsection{Reviewer F}

\quotereview{How is the needed information collected?}
We discuss the information that \sysname needs and how they are collected in \secref{s:sysname-overview}.

\quotereview{Is CLP the right metrics?}
Please refer to~\secref{rr:common3}.

\quotereview{Clarifying design}

We made numerous changes to clarify the design and, particularly, our model. 


%% file: abstract.tex
\noindent
{\bf Abstract--} 
  Cloud providers install mitigations to reduce the impact of network failures within their datacenters. Existing network mitigation systems rely on simple local criteria or global proxy metrics to determine the best action. In this paper, we show that we can support a broader range of actions and select more effective mitigations by directly optimizing end-to-end flow-level metrics and analyzing actions holistically. To achieve this, we develop novel techniques to quickly estimate the impact of different mitigations and rank them with high fidelity. Our results on incidents from a large cloud provider show orders of magnitude improvements in flow completion time and throughput. We also show our approach scales to large datacenters.

%% file: intro_ramesh2.tex
\section{Introduction}
\label{sec:intro}


Datacenter networks often incur a variety of (concurrent) failures ranging from failed or lossy links to localized, persistent congestion. Repairing these failures takes time~\cite{GCP-Incidents}. For example, operators require days to replace optical links and hours to fix hardware-induced packet corruptions~\cite{Wu_NetPilot, Bogle_TEAVAR, Gao-Scouts,Zhuo_CorrOpt}. As a result, cloud providers install mitigations to reduce the impact of failures while they work on repairs. 
In this paper, we focus on \textit{network-level} mitigations\footnote{We use network-level mitigations and mitigations interchangeably. We briefly discuss application-level mitigations in \secref{s:other-details}.} such as disabling links or switches and re-routing traffic. These are effective because of the path and resource diversity in datacenter networks.

Given the economic importance of cloud services and the increasing likelihood of failures at scale, it is essential to find and implement effective mitigations quickly. Today, cloud providers are increasingly using automation to mitigate each incident. For example, \azure uses automation for nearly $80\%$ of its incidents. An auto-mitigation system allows an operator to pre-define a limited set of potential mitigations for each type of failure. The system then chooses the ``best'' mitigation for each individual incident. At its core, an auto-mitigation system \textit{\textbf{ranks mitigations}} based on one or more criteria, and it must do so \textbf{\textit{quickly}} to be effective. At \azure, mitigations must be in place within $5$ minutes~\cite{Gao-Scouts} of failure localization.


\begin{table}[]
  \centering
  {\small
    \begin{tabular}{|l|c|c|c|c|c|c|c|}
        \hline
        Approach & Metric & E & G & U & B & S & P \\ \hline \hline
        NetPilot & Util/Drop & $\times$ & $\checkmark$ & $\times$ & $\checkmark$ & $\checkmark$ & $\times$\\
        CorrOpt & \#Paths & $\checkmark$ & $\checkmark$ & $\times$ & $\times$ & $\checkmark$ & $\times$\\
        Operator & \#Uplinks & $\times$ & $\times$ & $\times$ & $\checkmark$ & $\checkmark$ & $\times$ \\ \hline \hline
        \sysname & FCT/Tput & $\checkmark$ & $\checkmark$ & $\checkmark$ & $\checkmark$ & $\checkmark$ & $\checkmark$\\ 
        \hline
    \end{tabular}}
    \caption{\sysname is the only method that mitigates failures based on \textbf{E}nd-to-end {\bf G}lobal {\bf P}erformance metrics, considers {\bf U}ncertainty in future networking behaviors, supports a {\bf B}road range of actions and failure, and {\bf S}cales to large datacenters. ({\bf E}: End-to-End, {\bf G}: Global, {\bf U}: Uncertainty, {\bf B}: Broadly applicable, {\bf S}: Scalable, {\bf P}: based on Performance)}
    \label{tab:intro-compare-methods}
\end{table}

\azure uses \textit{local} criteria to assess mitigations.\footnote{Its 
auto-mitigation system uses local criteria to determine whether taking a fixed action is better than taking no action.} For example, disabling a link is acceptable if it leaves sufficient functional uplinks at the corresponding switch. The state of the art either uses {\it global proxy} metrics, such as the residual path diversity from the \cadd{top-of-rack (ToR)} switches to the spine of the datacenter~\cite{Zhuo_CorrOpt}, or \textit{global non-end-to-end} measures like packet loss and network utilization~\cite{Wu_NetPilot}. However, these methods can negatively impact customers by suggesting inadequate mitigations (\secref{sec:motivation}).



In this paper, we explore a new mitigation ranking criterion (\tabref{tab:intro-compare-methods}): the impact on the \textit{end-to-end} connection-level performance (CLP) metrics, throughput and flow completion time. We can quantify the \textit{global} impact using distributional measures of these quantities (averages and percentiles) across all connections in the datacenter. Failures adversely impact these global end-to-end metrics, and an ideal mitigation is the one that minimizes the impact. \cadd{For instance, if our goal is to optimize 1st percentile (1p) throughput, the best mitigation is the one with the least impact on 1p throughput (\secref{sec:design}).}

From a cloud operator's perspective, this criterion reflects the network performance that customers experience and is preferable to local or non-end-to-end metrics that may not correlate with customer-visible network performance. However, at the scale of modern datacenters, the feasibility of quickly ranking mitigations based on global end-to-end CLP measures is unclear. 

We introduce \sysname, a service for operators and auto-mitigation systems that quickly ranks mitigations while scaling to large clusters. \sysname leverages the insight that ranking mitigations only require an estimate of CLP distributions to produce an effective \textit{ordering}. Its \textit{CLP estimator} models traffic, routing, and transport behavior in sufficient detail to ensure ranking fidelity and, at the same time, produces results in just a matter of minutes.

To approximate CLP, \sysname must estimate \textit{per-flow} performance not just for the current network state (topology and routing) but for potential (unknown) future network states that may manifest while mitigation is in place. Flow performance can depend on other concurrent flows as well as flow arrivals and departures. The central challenge in \sysname is producing an estimate quickly while accounting for all these factors.

\begin{figure}
    \centering
    \includegraphics{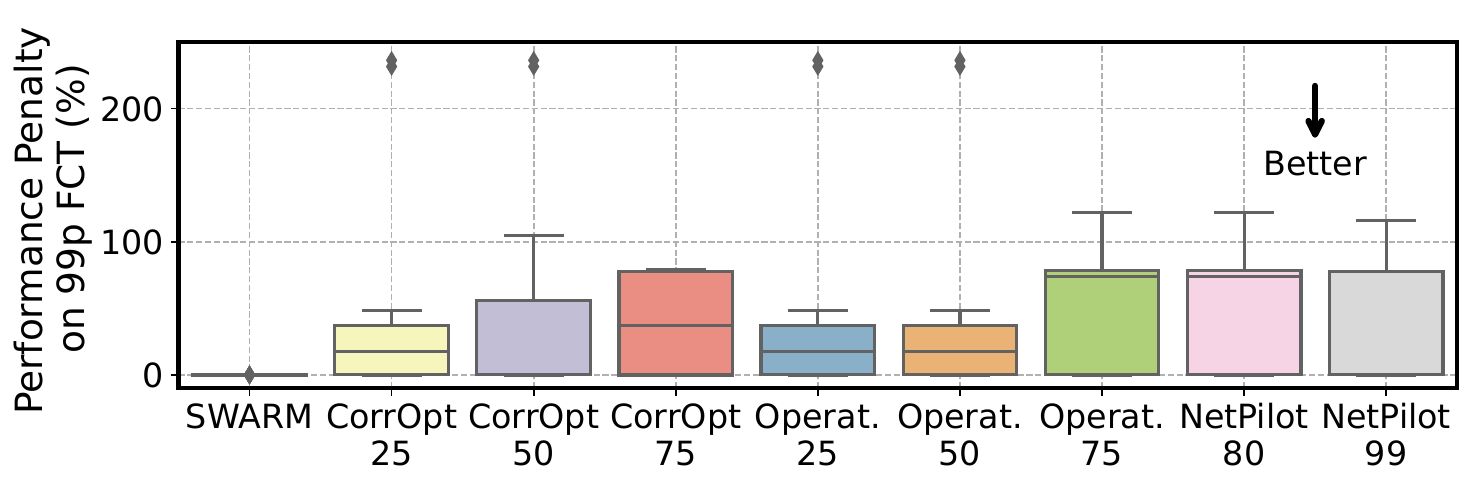}
    \caption{Our solution (\sysname) takes orders of magnitude better decisions by ranking mitigations using approximates of flow-level metrics and by accounting for uncertainties. The results are from scenario 1 \cadd{in Mininet}, see~\secref{ss:baseline-comparison} (Operat. = Operator).}
    \label{fig:mitigation-intro-ex}
\end{figure}

\vspace{.5mm}
\sysname overcomes these challenges as follows (\secref{sec:design}):
\begin{itemize}[leftmargin=*,nosep]
\item It takes flow arrivals, sizes, and communication probabilities as input distributions. It then samples a set of flow-level traffic traces to ensure statistical significance. For each demand, it generates routing samples to capture uncertainty in flows' paths. \sysname estimates CLP for each sample and then combines these estimates to create a composite distribution that succinctly captures traffic and routing variability. It uses this distribution to rank mitigations.

\item \sysname estimates CLP separately for long and short flows. Estimating CLP for short flows is easier since they experience less time-varying network behavior. \sysname takes care in modeling long flows along two dimensions: (1) whether their throughput is limited by loss or contention, and (2) how this limitation changes over time as flows arrive/depart and network bottlenecks shift.

\item It uses a suite of aggressive scaling methods, which include pipelining, parallelism, topology downscaling, and careful data structure design to estimate CLP quickly.
\end{itemize}


\vspace{.5mm}
Using CLP estimates allows \sysname to explicitly account for failure characteristics (\eg packet drop rate), reason about a broader range of mitigations (taking no action or bringing back a previously disabled link), and model failures (\eg packet drops below the ToR) that previous methods~\cite{Wu_NetPilot,Zhuo_CorrOpt} cannot (see \tabref{tab:intro-compare-methods}).




In summary, we make the following contributions:
\begin{itemize}[nosep,leftmargin=*]
\item We propose CLP-aware failure mitigation, which finds the mitigation with the least impact on network performance. This is a significant departure from state-of-the-art.

\item We identify sufficient approximations that allow us to build a robust and scalable CLP estimator that helps rank mitigations effectively (\figref{fig:mitigation-intro-ex}) and support a broader range of failures and mitigations compared to prior work (\tabref{tab:intro-compare-methods}).

\item We show \sysname is fast at scale and useful. For common failure scenarios (Scenarios 1 and 2 in \secref{sec:evaluation}), it picks either the best mitigation or one that is at most $9$\% worse \cadd{than the best mitigation}. It also outperforms the state-of-the-art by orders of magnitude. In more complicated failure cases which many prior work~\cite{Zhuo_CorrOpt,Wu_NetPilot} do not support, \sysname picks an action that is only $29\%$ worse than the best mitigation. 
\end{itemize}


%% file: case_for.tex
\section{CLP-Aware Mitigation}
\label{sec:motivation}

\newcommand{\fcs}{\textsc{fcs}\xspace}
\newcommand{\lcut}{\textsc{link cut}\xspace}
\newcommand{\lfcs}{\textsc{low fcs}\xspace}
\newcommand{\hfcs}{\textsc{high fcs}\xspace}

\begin{figure}[t]
  \centering
  \includegraphics[width=1.0\linewidth]{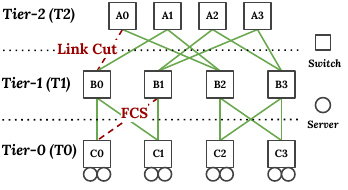}
  \caption{An example of two consecutive failures. First, the link between C0 and B1 experiences frame check sequence (FCS) errors. After mitigating, but before fixing this failure, a fiber cut between A0 and B0 causes congestion-induced packet drops.}
  \label{fig:example_two_drops_motiv}
\end{figure}

We use simplified versions of real-world incidents at \azure (\figref{fig:example_two_drops_motiv}) to explain the limitations of state-of-art auto-mitigation techniques and to illustrate the benefit of CLP-based impact assessments.

\vspace{.5mm}
\parab{Failure scenario.} Multiple link failures are common in cloud providers~\cite{Arzani-007, omnimon,Roy-NSDI17}. We emulate this in the Clos topology in \figref{fig:example_two_drops_motiv}. First, frame check sequence (\fcs) errors~\cite{Zhuo_CorrOpt} appear on a link. Operators mitigate this failure, but before they can physically replace the link, a fiber cut on another link (\lcut) causes congestion and packet loss.



We emulate (in Mininet~\cite{lantz2010network}, details in \secref{sec:evaluation}) a sequence of flow arrivals and successively apply each mitigation (or a combination of them) for \fcs and \lcut. \cadd{We use {\hfcs} and {\lfcs} to denote the drop rate of $\sim 5\%$ and $0.005\%$ respectively.} For this example, our goal is to maximize the 1st percentile (1p) throughput \footnote{\sysname can optimize quantiles of both throughput and FCT.}. We show how using \azure's troubleshooting guide, CorrOpt~\cite{Zhuo_CorrOpt}, and NetPilot~\cite{Wu_NetPilot} lead to substantial and unnecessary performance degradation.

\parab{Troubleshooting guides} in \azure disable any failed link (with drop rate $\ge 10^{-6}$) if at least half of the switch uplinks are healthy. This mitigation for {\fcs} achieves a 1p throughput of \underline{3.6 Mbps} and is optimal when the drop rates are high (\hfcs). However, it is conservative and static, and ignores the failure pattern, link location, and traffic demand. For example, leaving the lossy link in place in {\lfcs} and taking no action has a higher 1p throughput (\underline{14.2 Mbps}), while disabling the link causes congestion and impacts tail performance.
For {\lcut}, \azure's guidelines do not do anything in the face of congestion, which results in a 1p throughput of \underline{2.7 Mbps}. In this case, the 1p throughput is higher if we adjust WCMP~\cite{wcmp} weights to reduce traffic on congested links (\underline{3.2 Mbps}).

\vspace{.5mm}
\parab{CorrOpt~\cite{Zhuo_CorrOpt}} disables the link in {\fcs} if there is sufficient path diversity. This is sub-optimal for the same reason discussed above: depending on the failure properties (\eg drop rate and location), taking no action may lead to a higher 1p throughput. CorrOpt focuses on FCS errors and does not consider congestion induced by capacity drops, like {\lcut}.

\parab{NetPilot~\cite{Wu_NetPilot}} always disables the faulty link to optimize one of its health metrics (loss rate). This may not be the best option. In a \lfcs scenario, disabling the link drops the 1p throughput from \underline{15 Mbps} to \underline{3.6 Mbps}.
After the {\lcut}, NetPilot disables the congested link or switch to avoid additional drops. This exacerbates the problem and reduces the 1p throughput to \underline{3.17 Mbps}. A better strategy is to undo the previous mitigation and re-enable the {\lfcs} link when {\lcut} occurs, which results in 1p throughput of \underline{14.2 Mbps}. 

\vspace{.5mm}
\parab{Takeaways.} 
While this is a simplified example, operators can negatively impact customers in practice and cause extended outages if they fail to find an effective mitigation~\cite{Azure-RCA}. Rules with static thresholds in troubleshooting guides cannot capture correct mitigations because CLP impact depends on traffic demands. Path diversity measures (as in CorrOpt) cannot capture customer impact since they do not account for the failure characteristics. Non-end-to-end metrics like packet loss or utilization (as in NetPilot) often suggest disabling links, which discounts better mitigations. In contrast, \sysname ranks mitigations based on impact on end-to-end global measures.





%% file: design_ramesh.tex
\section{\sysname Design}
\label{sec:design}

\cadd{Network operators seek to improve flow completion time (FCT) and throughput in their datacenters~\cite{Homa-Montazeri,Alizadeh-DCTCP,pfabric,ExpandTime,teh2020couder}. These CLP metrics evaluate the performance properties of importance for short and long flows. Moreover, operators express their objectives as distributional statistics over the entire datacenter (\eg tail or average performance).} In this section, we describe how \sysname ranks mitigations in terms of their impact on CLP metrics, throughput and FCT.

\begin{figure}[t]
  \centering
\includegraphics[width=.95\linewidth]{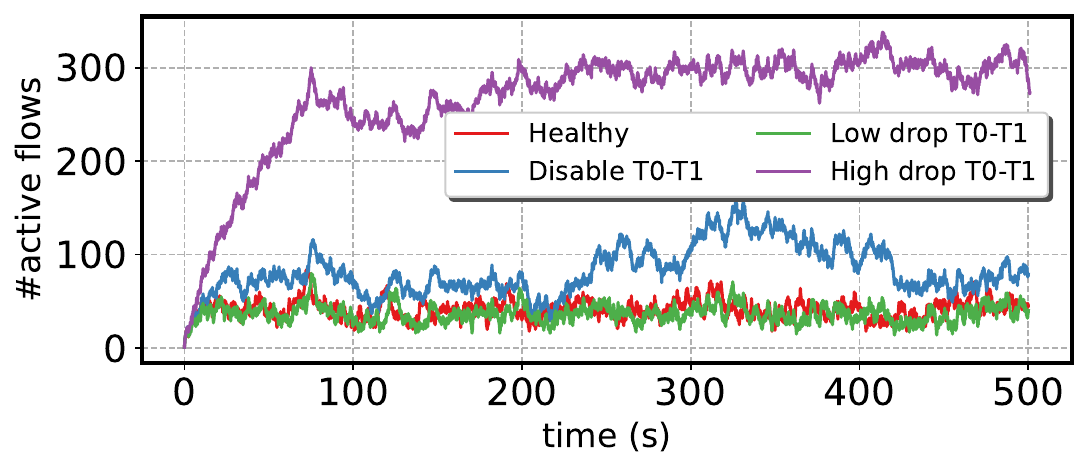}
  \caption{Failures and mitigations can increase flow durations, resulting in more active flows. (\figref{fig:example_two_drops_motiv} topology in \cadd{in Mininet})}
  \label{fig:example_num_active_long_flows}
\end{figure}

\subsection{Challenges and Insights}
\label{s:approach-challenges}

\noindent{\bf Challenges.} To optimize CLP objectives, \sysname must quickly estimate the distributions with sufficient accuracy to ensure effective mitigation ranking. This is hard:



\parae{Traffic characterization.} \sysname requires information about traffic demands to estimate CLP distributions for a given failure and mitigation. Instantaneous flow or ToR-level traffic matrices (TMs) can provide this information. However, fine-grained flow-level TMs are impractical to capture at datacenter scales and are sensitive to failures and mitigations. For example, packet drops often extend the flow durations, resulting in $3$~-~$4\times$ more concurrently active flows~(\figref{fig:example_num_active_long_flows}). ToR-to-ToR TMs, as in NetPilot~\cite{Wu_NetPilot}, are too ambiguous since they aggregate flows with different characteristics (\eg capacity drops impact long flows more than short flows).


\parae{Routing} determines the contention level at each link, which impacts throughput and FCTs. It depends on ECMP hash functions, as well as any existing failures in the network. The ECMP hash functions can change when links fail or switches reboot~\cite{sarykalin2008var}. \sysname must consider these factors.

\parae{Transport behavior.} \sysname needs to model transport behavior since CLP measures depend on congestion control algorithms (\eg BBR vs. Cubic), their parameters (\eg initial window size), and their reaction to failures (\eg packet drops). However, it is hard to model transport behavior while maintaining scalability. Accurate simulations over-index on specific protocols and are slow~\cite{henderson2008network}.
Faster approximate simulators do not account for lossy links~\cite{parsimon, mimicnet} or require a prohibitive amount of compute~\cite{mimicnet}. Existing formal models have limitations (see~\S\ref{seC:related_work}). For instance, fluid models~\cite{Kelly-TCP} capture steady-state for long flows, but datacenter flows are often short and do not reach a steady state~\cite{Homa-Montazeri}. 

\parae{Temporal and spatial dependencies.} CLP measures depend on the time-varying number of flows competing for bandwidth at a link. They also depend on where and for how long these flows experience congestion. For example, a flow bottlenecked at a link does not need its full fair share at other links. These bottlenecks may shift frequently due to flow arrivals and departures, and \sysname must capture them. 

{\small
\begin{table*}[t]
\centering
\small
\begin{tabular}{ l l l}
\textbf{Failure} & \textbf{Mitigation} & \textbf{Works that consider these failures/actions}\\
\hline 
\hline
\multirow{4}{*}{\textbf{Packet drop above the ToR}} & Take down the switch or link & NetPilot, CorrOpt, Operators \\
& Bringing back less faulty links to add capacity & $\times$ \\
& Changing WCMP weights & $\times$ \\
& Do not apply any mitigation & $\times$ \\
\midrule
\multirow{3}{*}{\textbf{Packet drop at ToR}} 
& Disable the ToR & Operators \\ 
& Move traffic \eg by changing VM placement & $\times$ \\
& Do not apply any mitigation & $\times$\\
\midrule
\multirow{5}{*}{\textbf{Congestion above the ToR}} & Disable the link. & NetPilot, Operators \\
& Disable the device & NetPilot, Operators \\
&  Bring back less faulty links to add capacity & $\times$ \\
& Change WCMP weights & $\times$ \\
& Do not apply any mitigation & $\times$ \\
\hline
\end{tabular}
\caption{List of failures and mitigations in \sysname. Disabling a device is a common mitigation for congestion~\cite{Wu_NetPilot}, see \secref{a:logic-mitigation}.\label{Table::failurelist}}
\end{table*}
}

\begin{figure}[t]
  \centering
  \includegraphics[width=1\linewidth]{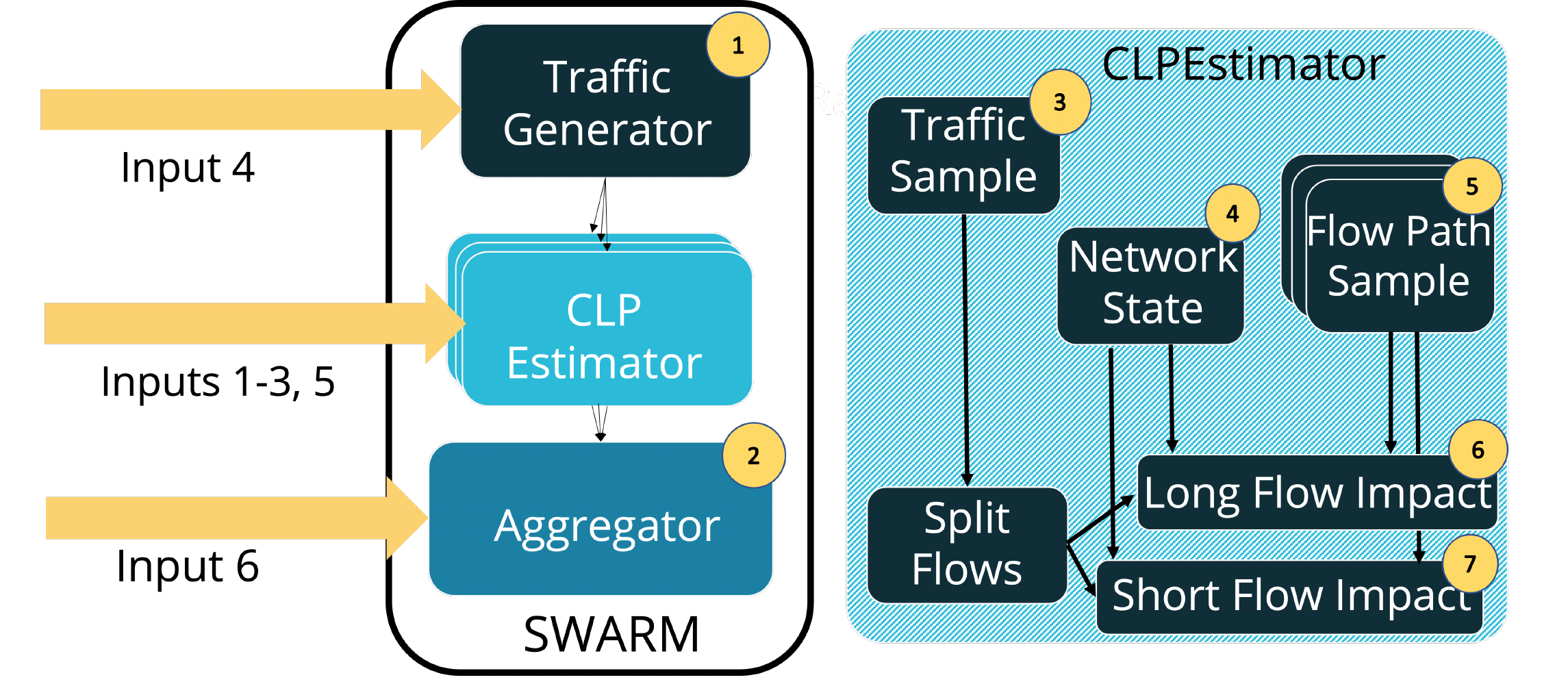}
  \caption{\sysname Design.}
  \label{fig:architecture}
\end{figure}

\parab{Approach.} At the core of \sysname (\figref{fig:architecture}) is a \textit{\impact} that estimates the {\em distribution} of throughput and flow completion time (FCT) for a given network, failure pattern, and mitigation set.

\sysname avoids the limitations of fine-grained flow-level TMs and uses an {\em approximate} TM distribution (\secref{s:sysname-overview}). It generates multiple (\circled{1}) TM samples (\circled{3}) from three inputs that cloud providers like \azure already collect: the flow arrival rates, the flow size distributions~\cite{Homa-Montazeri}, and the probability of server-to-server communication~\cite{privateeye}.

\sysname addresses routing uncertainty by evaluating CLPs on enough flow path samples (\circled{5}) to reach a target statistical confidence (\secref{sec:swarm:internals}). It generates these samples based on the network state (\circled{4}).

The trickiest challenge is to model the impact of losses and dependencies between concurrent flows on throughput and FCT. To this end, \sysname uses three techniques.

\parae{Epoch-based flow rate estimator.} \sysname uses a fast, scalable epoch-based flow rate estimator to handle temporal bandwidth changes and flow dependencies. It divides time into multiple epochs, recomputes CLPs in each epoch, and {\em combines} the results to find an overall estimate.

\parae{Traffic Classification.} \sysname needs to quickly find effective mitigations at scale rather than accurately estimating the flow performance, which can be slow~\cite{henderson2008network}. Therefore, \sysname divides traffic into \cadd{long (\circled{6}) and short flows (\circled{7})} based on their sizes~\cite{Alizadeh-DCTCP,Homa-Montazeri}. In each epoch, it estimates throughput and flow completion time separately. This approach results in significantly better mitigations (\secref{sec:evaluation}) for three reasons:

First,  failures affect short and long flow differently. Longer flows are exposed to network variations, while shorter flows only experience a snapshot of the network and have more predictable FCTs.

Second, congestion control algorithms~\cite{bbr,cubic} typically have a start-up phase to find available bandwidth. Short flows with few packets may finish during this phase. Therefore, they experience queueing delays caused by switch buffer occupancies rather than bandwidth limits~\cite{Alizadeh-DCTCP,Homa-Montazeri}. Accounting for these factors improves the effectiveness of our approach.

Third, previous studies~\cite{Alizadeh-DCTCP,Homa-Montazeri,pfabric} have shown that datacenter traffic is a mixture of short and long flows. Short flows are delay-sensitive, while long flows are throughput-sensitive. \sysname reports these separately and allows operators to adjust their mitigations based on their requirements. For example, they can prioritize short flows if their workload mainly consists of latency-sensitive short flows. Modeling these two classes separately also helps with scalability. Future work can explore modeling more than two classes of flows, such as those that are not short enough to ignore startup behavior.

\parab{Transport protocol abstraction.} \sysname uses an approximate model of transport protocols that is effective and scalable: (1) it assumes long flows are TCP-friendly~\cite{Alipourfard_Janus,Ben-StatisticalBWSharing}, which means each long flow grabs a fair share of the bottleneck bandwidth in the absence of failures. (2) under failures and packet drops, \sysname determines if long flows are {\em capacity-} or {\em loss-limited}. For capacity-limited flows, it computes their fair share of bandwidth. For loss-limited flows, it estimates the bandwidth at which the control loop converges under loss. \sysname extends existing max-min fair algorithms~\cite{Lavanya-sPerc,max-min} to detect and estimate both simultaneously. (3) \sysname assumes short flows do not reach steady-state and are impacted by packet drops or queueing delays rather than the bandwidth limits.


\sysname aggregates distributions (\circled{2}) from each traffic and routing sample. Operators use these to rank mitigations and set priorities based on one or more distributional metrics (\eg prioritize average throughput over FCT).

\subsection{\sysname: Inputs and Outputs}
\label{s:sysname-overview}

\noindent{\bf Inputs.} Operators or auto-mitigation tools can invoke \sysname
with the following inputs:

\begin{enumerate}[nosep,leftmargin=*]
\item Datacenter topology.
\item List of ongoing mitigations (if any).
\item Failure pattern (\eg \textit{estimated} loss rate) and location.
\item Data center traffic details (\eg TMs distributions).
\item Candidate mitigations to evaluate.
\item A \textit{comparator} that ranks mitigations by CLP estimates.
\end{enumerate}

\parae{Inputs 1-2-3.} Cloud providers use monitoring systems~\cite{snmp,syslog} and automated watchdogs~\cite{Gao-Scouts} to detect incidents and use different techniques~\cite{Wu_NetPilot,deepview,Arzani-NetPoirot} to localize the failure. They then create incident reports~\cite{Gao-Scouts,GCP-Incidents,Azure-RCA} that contain details of the incident. On-call engineers or automation systems use these reports to install mitigations that are active until the operator finds the root cause and repairs the failure. These reports contain the information \sysname needs, such as the failure location or pattern. The failure characterization (\eg packet drop rate) is sometimes imperfect, and operators may not be able to accurately localize the failure. \sysname can tolerate errors in packet drop rate (\secref{sec:evaluation}), and operators can incorporate the probability of different locations (\secref{sec::discussion}) and iteratively refine mitigations to deal with imperfect localization.

\parae{Input 4.}
\sysname requires simple characterizations of datacenter traffic: the flow arrival distribution, the server-to-server communication probability, and the flow size distributions. From these, it extracts a set of flow-level demand matrices (\secref{sec:swarm:internals}). These probabilistic characterizations of the inputs allows \sysname to be robust to traffic variability and ensure a desired level of statistical confidence in its estimation. 

\parae{Input 5.} \sysname requires a mapping from the failure types to a list of mitigations or a combination of mitigations. Cloud providers~\cite{autotsg,tsg} already document possible mitigations for each type of incident in their troubleshooting guidelines and can use these documents to create the mapping. Certain actions may require additional information, such as VM placements or WCMP weights. We assume operators use existing techniques~\cite{SuchaOptimalOblivious, VBP} to find values for these inputs. \tabref{Table::failurelist} shows a sample of failures and associated mitigations.


\parae{Input 6}. Operators can customize the \textit{comparator}. We currently support two types of comparators, and we can easily extend to others (see~\S\ref{sec:evaluation}). The \textit{priority} comparator considers throughput-based and FCT-based metrics in a pre-specified priority order. The \textit{linear} comparator is a linear combination of two or more of these metrics, where the operator specifies the weights. This flexibility enables operators to adjust their mitigation strategy based on their workloads. For example, if delay-sensitive short flows~\cite{Alizadeh-DCTCP,Homa-Montazeri} are the dominant workload, they can choose to prioritize the impact on short flows.

\parab{Outputs.} \sysname outputs the mitigation (or mitigation combination) with minimal impact as ranked by the comparator.


\subsection{\sysname: Internals}
\label{sec:swarm:internals}

\begin{figure}[t]
    \centering
    \includegraphics[width=0.8\linewidth]{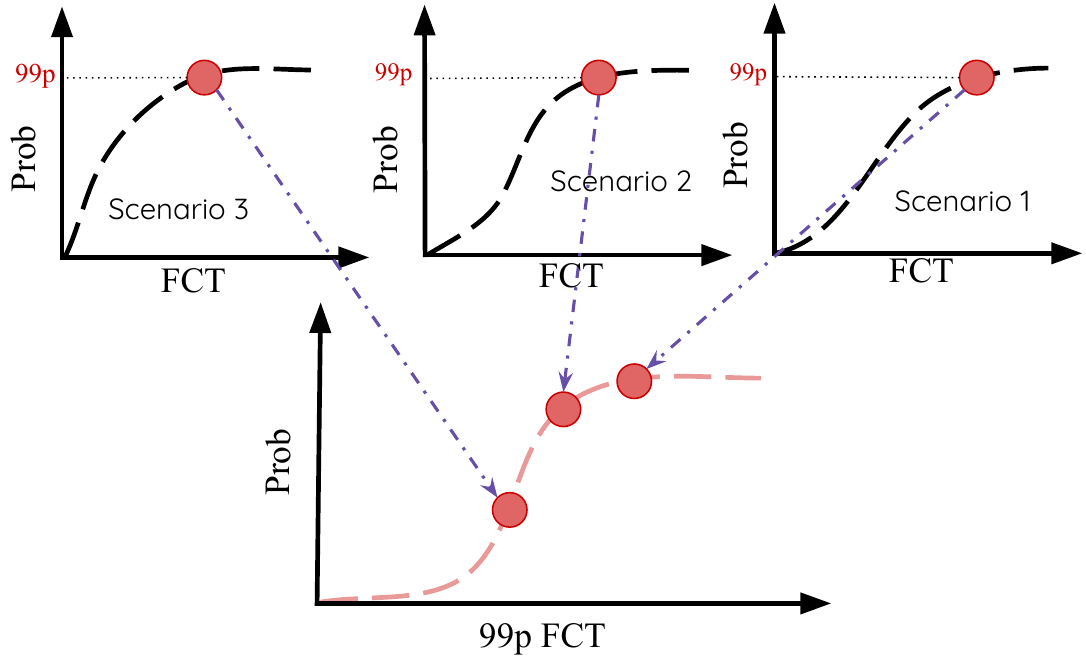}
    \caption{The composite distribution of 99p FCT obtained from the FCT distributions of different traffic and routing samples.}
    \label{fig:distribution-distribution}
\end{figure}

\noindent{\bf The \impact (\figref{fig:architecture})} takes a demand matrix $\traffic$ and a mitigation $\mitigationAction$ as inputs and estimates: (a) the distribution of average throughput across all the long flows in $\traffic$ and (b) the distribution of FCT across all the short flows in $\traffic$. \sysname can compute average throughput from FCT and vice versa using $\text{FCT} = \frac{\text{flow size}}{\text{throughput}}$ if necessary. \sysname calls the \impact for each candidate mitigation and ranks these mitigations based on the estimates (see \algoref{Alg:risk-main-function}).

\parab{How \sysname uses \impact.} 
\sysname samples $K$ different demand matrices (\circled{1} in \figref{fig:architecture}) and invokes the \impact to evaluate a given mitigation on each of them. The \impact then internally generates $N$ different \textit{routing samples} (\circled{5}), where each sample describes the path for every flow in the demand matrix. We choose $K$ and $N$ to ensure a desired confidence level (see below). 



Depending on the comparator, \sysname estimates the {\em distribution} of the {\em percentiles} of throughput and FCT across the traffic and routing samples (\figref{fig:distribution-distribution}). For instance, if the comparator uses 99p FCT, \sysname would extract the 99p from FCT distributions of the $N \times K$ samples and form a {\em composite distribution} of the 99p FCTs. The variance of this distribution captures the uncertainty in our estimates, which we can reduce by increasing the number of samples~\cite{samples} (\figref{fig:variance-flow-arrival}).

\sysname uses the composite distribution (\circled{2}) to compare and rank mitigations. This approach allows it to handle uncertainty explicitly and provide robust mitigation rankings.


\vspace{.5mm}
\parab{Network state representation.} \sysname models the network state (\circled{4}) using a graph $\topology = (\switchset, \linkset)$, where each edge $e$ has a capacity and a drop rate (0\% = healthy and 100\% = down), each node $v$ has a drop rate and a routing table, and each server $s$ maps to a switch. Before each invocation, \sysname \textit{updates} this state to reflect the mitigation (line 2 in \algoref{Alg:risk-main-function}). It uses data structures to ensure this step is efficient (\secref{s:other-details}).

\begin{figure}
    \centering
    \includegraphics[width=1\linewidth]{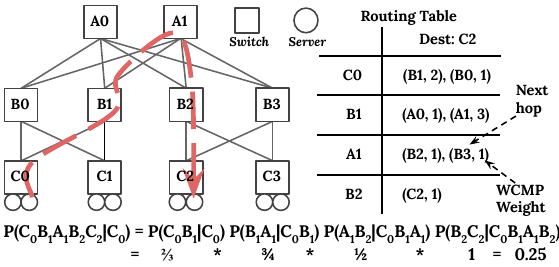}
    \caption{The probability of a flow taking a particular path.}
    \label{fig:routing-prob}
\end{figure}

\vspace{.5mm}
\parab{Modeling traffic variability.} The demand matrix $\traffic$ includes the arrival times, flow sizes, and their corresponding source and destinations. To create $\traffic$, \sysname uses input 4 (\circled{1}): for each flow, it randomly samples the arrival time from the flow arrival distribution, the source-destination pair from the server-to-server communication probabilities, and the flow size from the size distribution. \sysname then invokes \impact with $K$ demand matrix samples. It uses the \cadd{Dvoretzky–Kiefer–Wolfowitz (DKW)} inequality~\cite{DKWInequality} to determine $K$ based on a given confidence level $\confidenceLevel$. DKW provides confidence for the difference between an empirically sampled distribution and the ground truth distribution.

\parab{Modeling routing uncertainty.} The \impact handles routing uncertainty by generating $N$ different flow path samples (\circled{5}), each representing a different routing of the flows (\secref{s:approach-challenges}). First, it uses the DKW inequality~\cite{DKWInequality} to determine the $N$ to achieve a confidence level of $\confidenceLevel$. Then, it samples from the distribution of possible paths between source-destination pairs. It computes the probability of a specific path for a given source-destination pair based on routing tables and associated WCMP weights at each node (\figref{fig:routing-prob}).

\parab{Modeling the throughput of long flows.} \sysname models long and short flows separately (\lineref{6th:line:long-flow} and \lineref{7th:line:short-flow} in \algoref{Alg:risk-main-function}). Long flows typically reach their steady state, and their throughput depends on network variations and packet drops. 

\parae{Varying network conditions.} \sysname approximates network variations and the arrival and departure of flows that compete on a link by dividing time into discrete epochs (\algoref{Alg:Impact-Long-Flows}). It assumes stable conditions within each epoch (no flow arrival/departure) but allows for variations across them. At the beginning of each epoch, \sysname adds the newly arrived flows to the set of active flows (\lineref{6th:Line:add-new}). Then, it computes their bandwidth share (\lineref{7th:Line:compute-thru}). At the end of each epoch, \sysname updates the number of transmitted bytes for each flow, removes the completed flows, and records their overall throughput estimates (\lineref{8th:Line:record-start}-\lineref{15th:Line:record-end}).

\parae{Computing bandwidth share.} \sysname finds the rate of each flow within an epoch in two steps (\lineref{7th:Line:compute-thru} in \algoref{Alg:Impact-Long-Flows}):
\begin{enumerate}[nosep,leftmargin=*]
\item It computes the loss-limited throughput of each flow, as described below.
\item It executes a demand-aware extension of the water-filling approach~\cite{Lavanya-sPerc}, which uses the loss-limited throughput estimates as an upper bound on the flow throughput and provably converges in O({number of edges}). We have developed this extension (see ~\secref{sec:waterfilling}).
\end{enumerate}

\RestyleAlgo{ruled}
\begin{algorithm}[t]
{
\setstretch{1.25}
\small
\DontPrintSemicolon
\LinesNumbered
\caption{Impact on Long Flows.}
\label{Alg:Impact-Long-Flows} 
\KwIn{$\topology$. current network state.}
\KwIn{$\traffic$ = \{<source, destination, size, start time>\}.}
\KwIn{$\mathcal{R}$. the sampled routing of each flow.}
\KwIn{$\mathcal{I}$. measurement interval.}
\KwIn{$\epochsize$. epoch size.}
\KwOut{$\impactFlows_{l}$. distribution of throughput of long flows}
$\listcurrentflows \gets \{\}$ \;
$time \gets 0$\;
$\impactFlows_{l} \gets []$\;
\While{$\exists\; f \in \traffic:~f.\text{start} \ge time \; or \;  \listcurrentflows \neq \varnothing$}{\nllabel{4th:Line:epoch-start}
    $time \gets time + \epochsize$\;
    $\listcurrentflows.\text{add}(\{f\in \traffic:\; time - \epochsize \leq f.start < time\})$\nllabel{6th:Line:add-new}\; 
    $\throughput_f \gets \text{compute\_throughput}(\topology, \listcurrentflows, \mathcal{R})$ \nllabel{7th:Line:compute-thru}\;
    \For{$\text{flow}~f~\text{in}~\listcurrentflows$}{ \nllabel{8th:Line:record-start}
        $f.sent \gets \min(f.sent + \epochsize\throughput_f,~f.size)$ \;
        \If{$f.sent = f.size$}{
            $\listcurrentflows.\text{remove}(f)$\;
                \If{$f \in \mathcal{I}$}{$\impactFlows_{l}.\text{append}(\frac{f.size}{f.dur})$\;}
        }
    } \nllabel{15th:Line:record-end}
}\nllabel{17th:Line:epoch-start}
\Return{$\impactFlows_{l}$}
}
\end{algorithm}

\parae{Modeling loss-limited throughputs.} It is possible to compute the throughput under packet drop analytically, but these models are specific to certain congestion control protocols and cannot easily extend to other variants (\eg~\cite{Alizadeh-DCTCP}). \sysname overcomes this limitation by using an empirically-driven distribution of the loss-limited throughputs. To find this distribution, \sysname measures the average throughput of a long flow under different network conditions (\eg drop rate, latency) through experiments in a small testbed. In each experiment, it ensures that link capacities are high enough so that they never become bottlenecks, and the drop rate is the only limiting factor. \sysname repeats the experiment multiple times for each network condition to create a robust distribution~\cite{DKWInequality}. Depending on the uncertainty in transport protocols running in the data center, we can also change the mix of transport protocols we use. \sysname then uses this distribution to sample the drop-limited throughputs. See~\S\ref{appendix:offline-measurement} for details.

\parab{Modeling the FCT of short flows.} Prior work~\cite{Mellia-tcpmodel-shortflows} develops an analytical model for the average FCT across flows. We are unaware of any models that can estimate the distribution. In developing \sysname, we observe that short flow are more predictable since they do not stay in the network long enough to be affected by network variations. Thus, a simple empirical model is sufficient to estimate their FCT distribution.

\sysname finds a short flow's FCT distribution (\circled{6}) by estimating (a) its RTT and (b) the number of RTTs to deliver its demand. A short flow's FCT is equal to the average duration of RTTs multiplied by the required number of RTTs. We derive the distribution of each metric separately and then combine them to compute the FCT distribution. 

We compute the distribution of the number of RTTs to deliver a flow's demand by conducting offline experiments in a small testbed. We repeat the experiments for different configurations (flow size, slow start threshold, initial congestion window) and network settings (drop rates, RTT) and store the results in a table that maps the setting to the measured distribution.



Next, we estimate the RTT, which is equal to the sum of the propagation delay (a constant determined by the flow's path) and the queueing delay along the flow's path. To estimate queueing delay, we collect data from sending small flows on links with different utilization and active flow counts (\S\ref{appendix:queueing}).

\sysname combines these distributions to derive the  FCT distribution for short flows. It multiples the number of RTTs by the sum of the propagation delay and queueing delay.

\subsection{Expressivity, Scaling, and Robustness} 
\label{s:other-details}

\noindent{\bf Expressivity.} \sysname supports any failure or mitigation as long as we can model it as changes to the network state or the traffic. \sysname does not need the root cause of a failure or details of a mitigation but only needs to understand their observable impact (\eg packet drop, port down). This flexibility allows \sysname to support a wide range of failures and mitigations (\tabref{Table::failurelist}). Existing production and state-of-the-art systems ~\cite{Wu_NetPilot, Zhuo_CorrOpt} do not support many of these failures, while cloud providers~\cite{Arzani-007,omnimon,Roy-NSDI17,Gill-UnderstandingFailures,Zhuo_CorrOpt,Wu_NetPilot}) commonly observe them.

\parab{Robustness.} \sysname's network-level mitigations should ideally mask network failures. However, this masking is not always perfect, and a cloud service might react to a mitigation (or the failure in general) by changing the traffic demand (\eg using retries). For this reason, \sysname does not model a fixed traffic demand. Instead, it draws multiple traffic samples from the historical distribution of flow arrivals, sizes, and communication probabilities to ensure statistical significance (\secref{sec:swarm:internals}). \cadd{When operators do not have such statistics
(\eg after a previously unseen failure or a data center expansion), \sysname uses a distribution that captures maximum uncertainty~\cite{robinson2008entropy}.}

\cadd{\sysname carefully models uncertainty, but its CLP estimates may not align with what operators observe after installing the mitigation. This can happen if an uncommon failure manifests itself that was not captured during the sampling process or if the traffic changes and the historical distribution is not representative anymore.} In such instances, the auto-mitigation system that uses \sysname must update its inputs and invoke \sysname again to revise the mitigation. In other words, mitigation does not have to be a single-shot process and can be adjusted over time, especially since failure diagnosis might take hours to days~\cite{Gao-Scouts, Wu_NetPilot}.

\parab{Scaling.}
\sysname must find a mitigation action quickly, even on large clusters. We scale it using the following techniques:

\parae{An ultra-fast max-min fair computation algorithm.}
We use an approximate computation of network-wide max-min fair share rates~\cite{max-min}, which provides significant speedup over the state-of-art methods~\cite{Lavanya-sPerc} without affecting quality.

\parae{Efficient network state and traffic update.} The traffic trace is independent of the network state. This means \sysname can compute the traffic samples offline. However, when \sysname invokes \impact, it must update network state to reflect each mitigation and then re-compute routing samples. To scale, \sysname separates the topology representation from the traffic representation. It models the former as a graph (\secref{sec:swarm:internals}) and sorts the latter into a list of tuples (source and destination server, flow size, and flow start time). This design enables, for example, disabling a link by changing the drop rate in $\topology$ to 100\% or updating the traffic if a mitigation such as VM migration modifies a flow.



\parae{Parallelism and pipelining.} \sysname (a) evaluates demand and routing samples in parallel, and (b) parallelizes and pipelines routing sample generation with epoch execution.

\parae{Reducing the number of epochs.} After these steps, the bottleneck is the number of epochs in \algoref{Alg:Impact-Long-Flows}. We use two techniques. Firstly, we initialize on an already warmed-up network instead of starting from an empty network. This eliminates the need for a set of epochs at the beginning to mimic the cold-start effect. Secondly, \cadd{the residual impact of flows that compete within an epoch diminishes over time. Therefore, epochs with large time differences present independent network snapshots. This observation allows us to compute their CLPs independently in parallel and combine the results to form an overall distribution.}

\parae{Traffic \cadd{downscaling}.} Following POP~\cite{Narayanan-POP}, \sysname \cadd{downscales} the demand matrix with minimal impact on throughput. It splits a network with link capacity $c$ into $k$ sub-networks with link capacity $\frac{c}{k}$ and divides traffic randomly across these sub-networks. POP~\cite{Narayanan-POP} recommends choosing a value for k that is much smaller than the number of flows, which allows each partition to capture the network contention. This approach works with any flow arrival distribution. We use Poisson distributions~\cite{BeyondFattrees,projector,ExpandTime}, where assigning flows randomly is the same as downscaling the arrival rate based on the Poisson splitting property.

%% file: evaluation.tex
\section{Evaluation}
\label{sec:evaluation}

Our prototype of \sysname has 1500 lines of Python.
We evaluate it on three categories of incidents and show it outperforms existing methods. We also show it scales and finds the best mitigation for a 16K-server topology within 5 minutes.

\subsection{Methodology}
\label{sec:eval-meth}

\noindent{\bf Metric.} We evaluate each approach by computing the \textit{Performance Penalty (\%)}, which is the relative difference between the CLP metrics that result from the best possible mitigation and the one each technique suggests.
This metric captures the unnecessary performance degradation caused by a technique choosing sub-optimal mitigation.
Often, the difference between the best mitigation and the ``runner-up'' is insignificant, resulting in a small penalty. In contrast, the difference between the best mitigation and the one the baselines choose can be as high as 200\%.

\parab{Experimentation setup.}
We evaluate \sysname using Mininet~\cite{lantz2010network}, NS3~\cite{henderson2008network}, and a physical testbed. Here is the summary of our setup (see~\secref{appendix:experiment-details} for more details):

\parae{Traffic characterization.} We use \azure production logs to derive the flow inter-arrival time, a commonly used distribution from DCTCP~\cite{Alizadeh-DCTCP} for our flow size distribution, and~\cite{hpcc} for server-to-server communication probability. We ensure that the trace duration is long enough to capture all flow sizes and that we do not capture an empty network's effect. Finally, we run on 30 different traces to ensure robustness.

\parae{Emulation Setup.} We mainly use Mininet for our evaluation since it leverages a real TCP/IP stack from the Linux kernel. We extend it to improve the fidelity of our results (monitoring systems, queueing disciplines). We report results on 57 scenarios (over 4000 hours of experiment) across 3 types of common incidents in cloud providers~\cite{Arzani-007,Roy-NSDI17,Zhuo_CorrOpt,Gill-UnderstandingFailures,Wu_NetPilot} (\tabref{Table::failurelist}). We describe these scenarios in \secref{ss:baseline-comparison} and \tabref{tab:experiment-details}. We use the Clos topology from Figure~\ref{fig:example_two_drops_motiv}.

We seek to emulate $1500$ flows arriving per second per server, which results in $12,000$ flows every second. These are on links with $40$ Gbps bandwidth and 50 \cadd{$\mu$s} propagation delay.
However, running Mininet on a VM with 64 cores and 256 \cadd{GB} cannot emulate this demand. Instead, we use~\cite{Konstantinos_downscale_simulation,Shrink-Pan} to downscale the traffic and link capacities by 120$\times$(see~\secref{a:exp-other-details}). We run each emulation for 500 s and measure the performance for flows that start within [50, 150) s to avoid capturing effects from an empty network. We report results for both Cubic~\cite{cubic} and BBR~\cite{bbr} (and DCTCP~\cite{Alizadeh-DCTCP} in our simulation).

\parae{Simulation Setup.} We use a Clos topology with 128 servers, 32 ToRs, 32 T1s, and 16 T2s, all connected by 20 \cadd{Gbps}, 100 \cadd{$\mu s$} links. We use DCTCP~\cite{Alizadeh-DCTCP} as the congestion control algorithm to show generality. Each of our traffic traces is 10 s long, and we measure the flows that start within [0.5, 1) s\footnote{\cadd{We adjust the duration based on the flow arrival rate and link bandwidths. When these values are smaller as in our Mininet experiments, it will take longer for a certain number of flows to arrive, and each flow remains in the network for a longer period. Thus, we need to cut a longer duration at the beginning to ensure we are not capturing the effect of the empty network.}}. We use DCTCP~\cite{Alizadeh-DCTCP} and FbHadoop~\cite{hadoop-workload} flow size distributions.


\parae{Testbed Setup.}
We use a different variant of Clos (see~\secref{a:setup}) with 32 servers, six TORs, four T1s, and two T2 switches, all connected by 10 \cadd{Gbps} 200 \cadd{$\mu s$} links.
All switches are Arista 7050QX-32.
We introduce random packet loss in the testbed using a user-defined \cadd{access control list} (ACL) to match bits on the IP ID field in packet headers and directly modify the Broadcom firmware to drop packets the ACL matches. Thus, packet loss rates in the testbed are powers of two based on the number of bits the ACL matches on. We evaluate the impact of the failure and each mitigating action on the testbed with a traffic load of 3000 flows per second. Each trace is 30 seconds long, and we measure across flows that start within [2, 5) s.

\parae{\sysname Parameters.} We use 32 different random traffic traces and 1000 routing samples based on~\secref{sec:swarm:internals}. For our baseline comparisons (with Mininet), each trace is 200 s long. We compute the performance over all the flows that start within [50, 150) s. Each epoch is 200 \cadd{ms}. We also consider any flow with a size $\leq$150 \cadd{KB} short.

\begin{figure*}[h]
  \centering
  \subfigure{\includegraphics[width=0.75\linewidth]{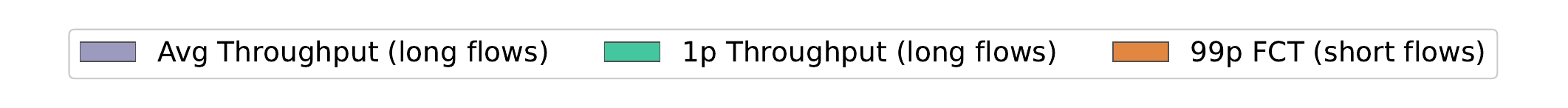}}
    \addtocounter{subfigure}{-1} \vspace{-4mm}\\
  \subfigure[PriorityFCT]{\includegraphics[width=.95\linewidth]{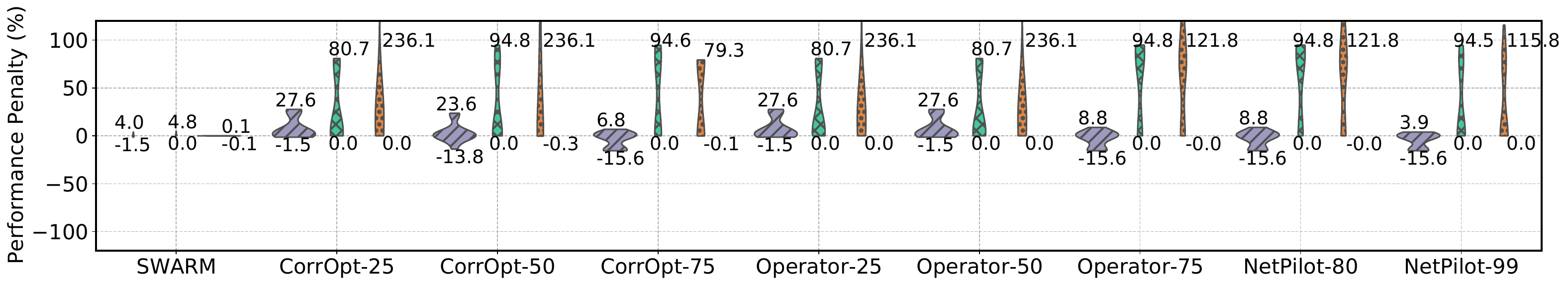}\label{fig:scenario_1_results:priorityfct}}
  \vspace{-1mm} \\ 
  \subfigure[PriorityAvgT]{\includegraphics[width=.95\linewidth]{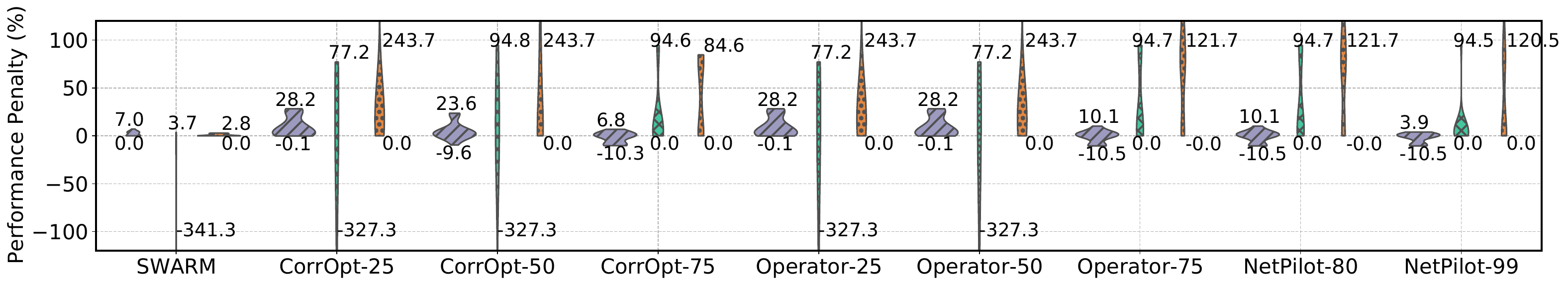}\label{fig:scenario_1_results:priorityavgt}}
  \caption{\textbf{Comparison of \sysname and other baselines under Scenario 1 \cadd{in Mininet}.}
  \sysname achieves 793$\times$ lower performance penalty on 99p FCT in the worst case compared to the next best baseline on PriorityFCT. \sysname is the only technique that achieves near-optimal performance across all three metrics and performs equally well across both comparators.  
  }
  \label{fig:scen1-all-baselines-comparison}
\end{figure*}

\parab{Baselines.} We compare \sysname to:
\begin{description}[nosep,wide]
\item[NetPilot~\cite{Wu_NetPilot}] iterates through each possible mitigation, computes the maximum link utilization, and picks the action that minimizes utilization. NetPilot does not model link utilization on faulty links, so it always disables corrupted links. We report these results as NetPilot-Orig.
We also extend NetPilot to mitigate only if the resulting maximum link utilization is below a threshold. We report these
results as NetPilot-80 for an 80\% and NetPilot-99 for a 99\% utilization threshold.

\item[CorrOpt~\cite{Zhuo_CorrOpt}] only considers link corruption failures. It disables a link if the number of remaining paths to the spine after taking the action is above a threshold. We consider three thresholds: CorrOpt-25, CorrOpt-50, and CorrOpt-75, which use a threshold of 25\%, 50\%, and 75\%, respectively.

\item[Operator playbooks.] When an FCS error occurs above the ToR where there is path
redundancy, the \azure playbook will disable the affected link if the number of remaining uplinks at the switch is above a certain threshold.
We consider three thresholds: Operator-25 uses a 25\% threshold, Operator-50 uses 50\%, and Operator-75 uses 75\%. When there is packet loss of more than $10^{-3}$ at or below the ToR, the playbook will drain the affected nodes, which is expensive and risks VM reboots or interrupts. Otherwise, it would take no action.
\end{description}

Some baselines cause a partitioned network under certain scenarios in part due to the smaller scale in our Mininet evaluations. Unless noted otherwise, we only report cases where all baselines keep the network connected for a fair comparison.

\parab{Comparators.} Most experiments use two priority comparators (\secref{s:sysname-overview}) (\secref{appendix:other-comparators} shows \sysname achieves low penalty across two other comparators including a linear one as well):

\begin{description}[nosep,wide]
\item[PriorityFCT] minimizes the 99p FCT. It uses two tiebreakers, 1p throughput followed by average throughput.

\item[PriorityAvgT] maximizes the average throughput first, using two tiebreakers, 99p FCT, followed by 1p throughput.
\end{description}

\vspace{.5mm}
Two mitigations are tied on a particular metric if they are within 10\% of each other on that metric.

\subsection{Baseline Comparisons}
\label{ss:baseline-comparison}

We evaluate \sysname over three different failure scenarios, which are common in production incidents at \azure.

\parab{Scenario 1: Link-level packet corruption with network redundancy.} In this scenario, we evaluate different combinations of two links consecutively experiencing FCS errors with a drop rate of \cadd{$\sim 5\%$} ({\em high}) or  \cadd{$\sim 0.005\%$} ({\em low}). This is the most common failure pattern at \azure, and both drop rates are detected and reported as incidents. In this case, the viable mitigations are doing nothing, disabling the link, undoing past mitigations, changing WCMP weights, or any feasible combination of these. All baselines support this scenario because the link failure is above the ToR, and there is path redundancy.



In \figref{fig:scen1-all-baselines-comparison}, we compare the performance penalty of \sysname to the baselines for both comparators. We do not compare to NetPilot-orig in this scenario since it partitions the network in 16 out of 32 failure pairs, and the results on the remaining failures are not statistically significant.
We use violin plots to show the performance penalties' distribution across all incidents for each candidate approach and each metric.
A tall violin plot indicates that performance penalties span a wide range, while a short and wide plot indicates that penalties are clustered within a small set of values. Even though the baselines do not explicitly use comparators, the best mitigation depends on the comparator, which causes the penalty to change for the baselines across different comparators.

\begin{figure}[t]
    \centering
    \includegraphics[width=1.0\columnwidth]{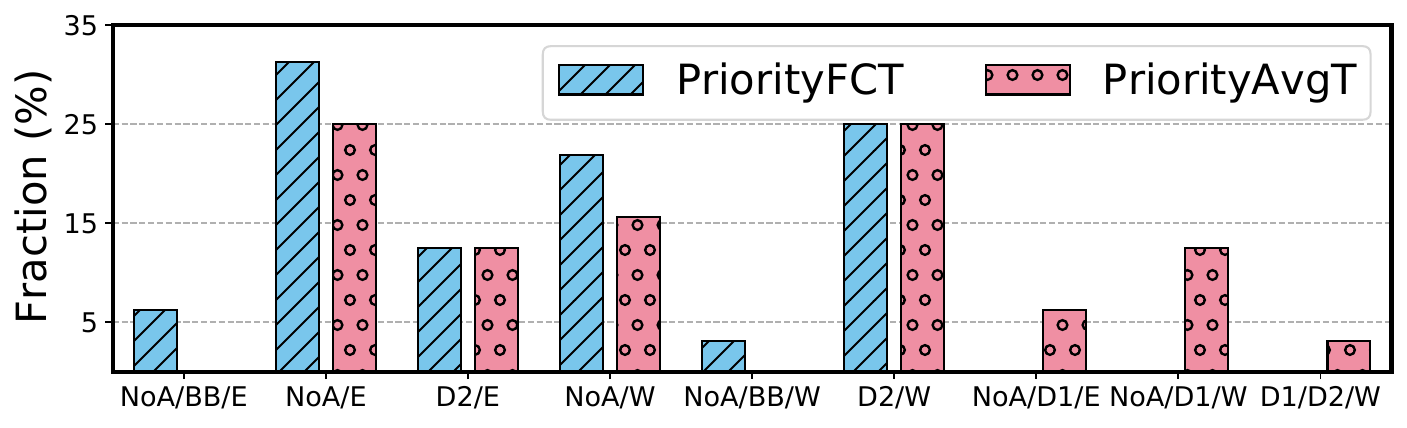}
    \caption{\textbf{\sysname's actions in Scenario 1 (see~\S\ref{sec:evaluation}).} It chooses from nine action combinations and decides to take no action in more than 25\% of the time. (NoA: No Action on link 2, D2: Disable link 2, BB: Bring Back link 1, D1: Disable link 1, W: WCMP Routing, E: ECMP Routing)}
    \label{fig:scenario_1_actions}
\end{figure}

\sysname outperforms all baselines with a performance penalty consistently close to zero. When using the Priority-FCT comparator, \sysname has a maximum FCT performance penalty of 0.1\%, compared to 79.3\% penalty of the closest baseline, CorrOpt-75 (\figref{fig:scenario_1_results:priorityfct}).
For the Priority-AvgT comparator, \sysname's maximum
penalty on average throughput is similar to several of the baselines, such as NetPilot-99 and CorrOpt-75. However, \sysname reduces the performance penalty across {\em all three} CLP metrics while the baselines suffer from high performance penalties across at least one (\figref{fig:scenario_1_results:priorityavgt}). \sysname's superior performance is due to its ability to analyze a broader set of mitigations and to choose one that reduces performance impact on all CLP metrics (see~\secref{a:benefit-two-ex}).

The negative penalty in some cases is because of the inherent trade-off between different metrics. For instance, the primary objective in~\figref{fig:scenario_1_results:priorityfct} is to minimize short flows' 99p FCT. The optimal mitigation is the one that has the best FCT, even if it worsens other metrics. For example, CorrOpt-75 selects actions that increase the 99p short flow's FCT (highest priority) by up to 80\% compared to the optimal action but achieve better average throughput (lower priority). 

We show the diversity of \sysname's proposed mitigation under each comparator in \figref{fig:scenario_1_actions}. We focus on the mitigation for the second failure since the action space is larger and includes options such as bringing back a previously disabled link. We find \sysname chooses from nine different possible mitigations and decides to take no action in more than 25\% of the cases. In two scenarios under PriorityFCT, it not only takes no action on the second failure but also \emph{reinstates} the faulty link it previously disabled from the first failure (action NoA/BB/E). \sysname considers the failure locations, their intensity, and the traffic demands and decides to preserve those links rather than eliminate their capacity and cause congestion. Under PriorityAvgT, \sysname chooses combinations of mitigations such as disabling both links and adjusting WCMP weights.

\begin{figure}[t]
    \centering
    \subfigure[PriorityFCT]{\label{fig:congestion_scenario:fct}\includegraphics[width=0.46\textwidth]{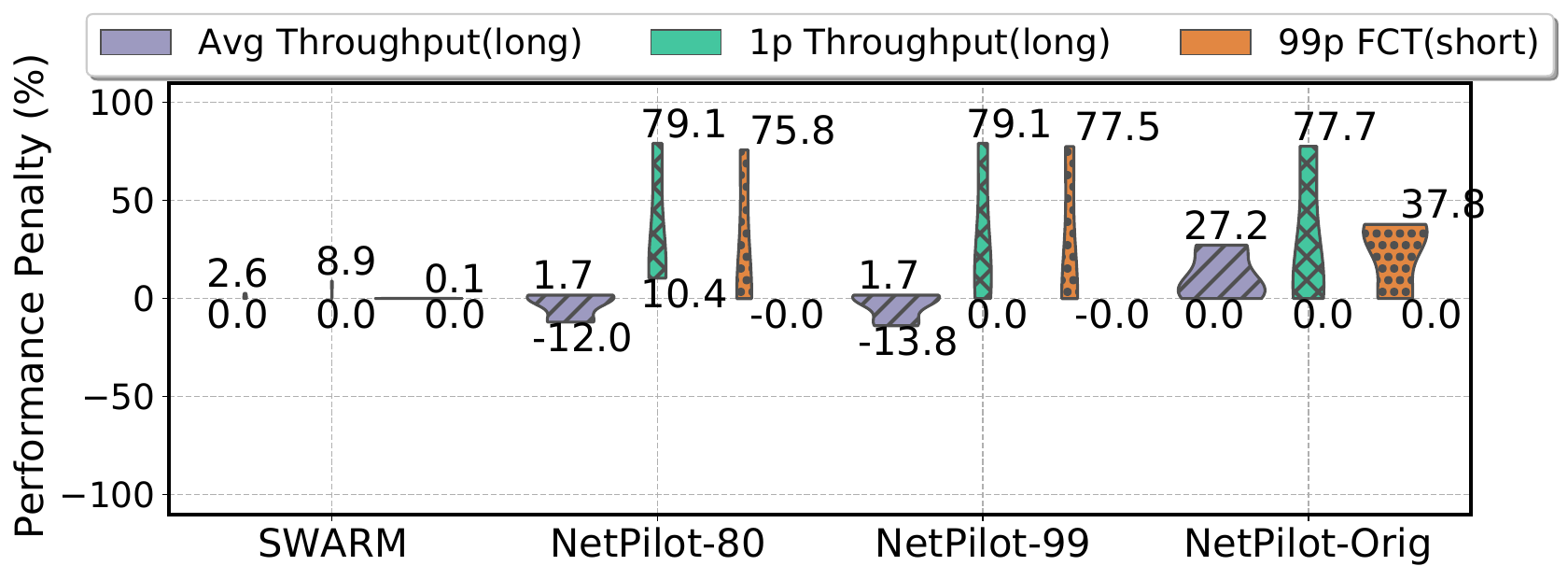}}
    \subfigure[PriorityAvgT]{\label{fig:congestion_scenario:avg}\includegraphics[width=0.46\textwidth]{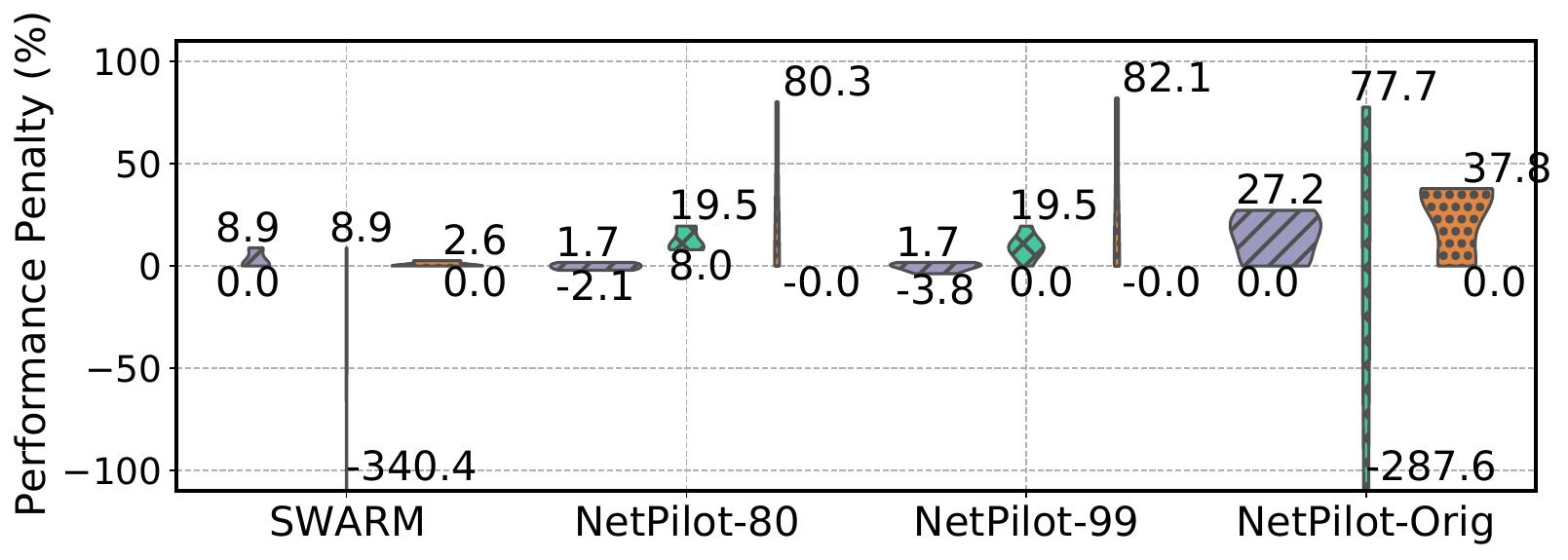}}
    \caption{\textbf{Comparison of \sysname and NetPilot variants under Scenario 2 \cadd{in Mininet}.} Under the PriorityFCT comparator, \sysname always chooses a mitigation with near-optimal FCT performance ($\leq$ 0.1\%), while the next best approach chooses mitigations up to 37.8\% worse than optimal. \sysname is also the only approach with low performance penalty across all three metrics.}
    \label{fig:congestion_scenario}
\end{figure}

 \begin{figure}[t]
    \centering
    \subfigure[Priority FCT]{\includegraphics[width=0.96\linewidth]{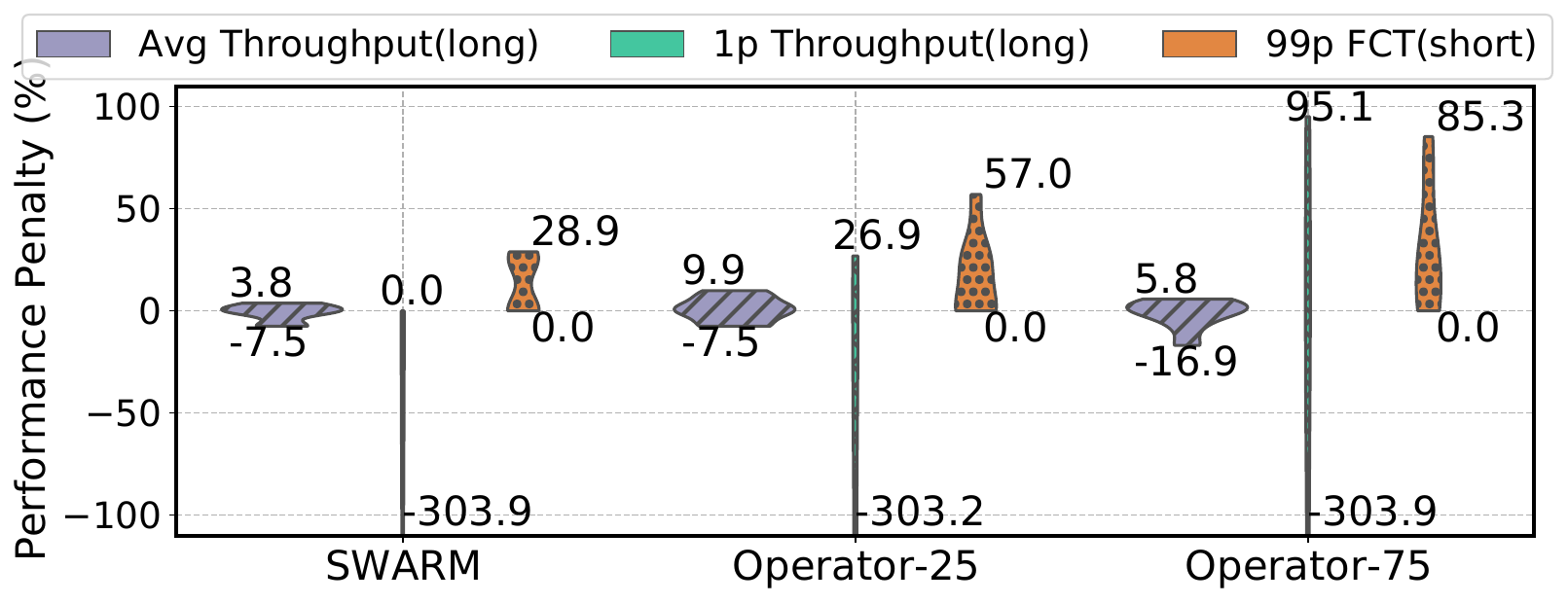}}
    \subfigure[Priority AvgT]{\includegraphics[width=0.96\linewidth]{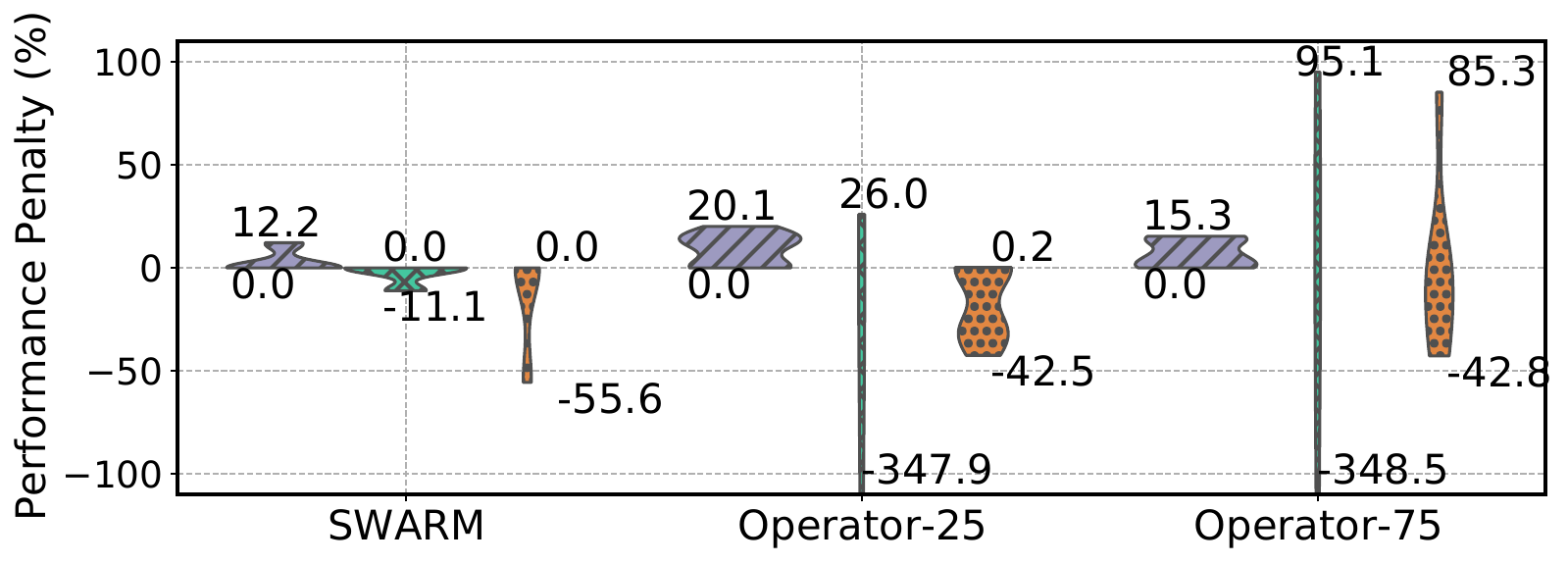}}
    \caption{\textbf{Comparison of \sysname and operator playbooks on Scenario 3 \cadd{in Mininet}.} Under PriorityFCT, \sysname has 2$\times$ lower 99p FCT penalty in the worst-case compared to operator playbooks. It also protects other non-priority metrics while optimizing for the priority metric (it outperforms baselines in both average throughput and 99p FCT under PriorityAvgT).}
    \label{fig:scenario-tor-drop}
\end{figure}


\parab{Scenario 2: Congestion on a link.} Operators shut down several faulty links (prior failures), causing the network to become over-subscribed. Meanwhile, an aggregation-core layer starts operating at half capacity (due to fiber cuts). CorrOpt and operator playbooks do not support this as they ignore traffic dynamics.
NetPilot can reason about congestion but assumes the rest of the network is under-utilized. We evaluate \sysname and NetPilot under two failure patterns: (i) where the network is under-utilized and (ii) where a second link drops packets and reduces network capacity (\tabref{tab:experiment-details}). 

\figref{fig:congestion_scenario} compares the performance penalty of \sysname to the NetPilot variants for both comparators across both failure patterns.
\sysname achieves consistently low penalty for its target CLP metric: 99p FCT in \figref{fig:congestion_scenario:fct} and average throughput in \figref{fig:congestion_scenario:avg}.
Since NetPilot assumes the rest of the network is under-utilized,
it aggressively disables links and causes a large performance impact. Under PriorityFCT, \sysname chooses
a mitigation with near-optimal performance on FCT, while the next best approach suffers an FCT penalty of 38\%.
Under the PriorityAvgT comparator, the NetPilot-80/99 variants result in lower impact on average throughput, but at the cost of increased performance penalty in at least one other metric (\eg NetPilot-80 achieves $7.2\%$ less penalty in terms of average throughput at the cost of $~80\%$ penalty on $99$p FCT) while \sysname is the only technique that performs well across both comparators and all three metrics.




\begin{figure*}[h]
    \centering
    \subfigure[Runtime for different topology sizes]{\includegraphics[width=0.30\linewidth]{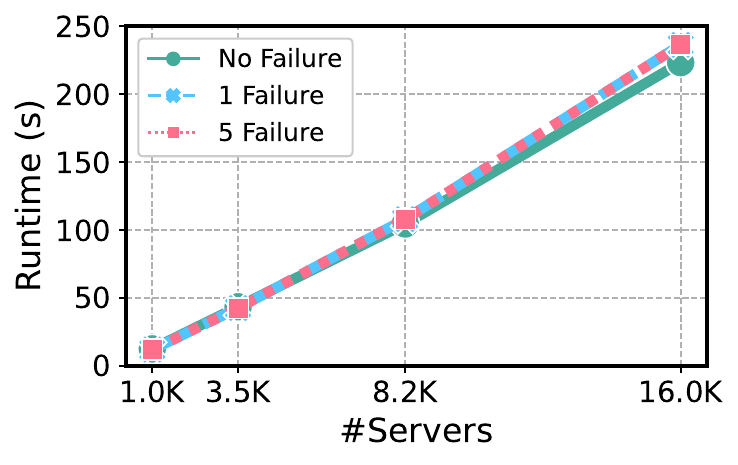} \label{fig:scalability:run-time}} 
    \hfill
    \subfigure[Estimation Error of different scalability choices]{\includegraphics[width=0.34\linewidth]{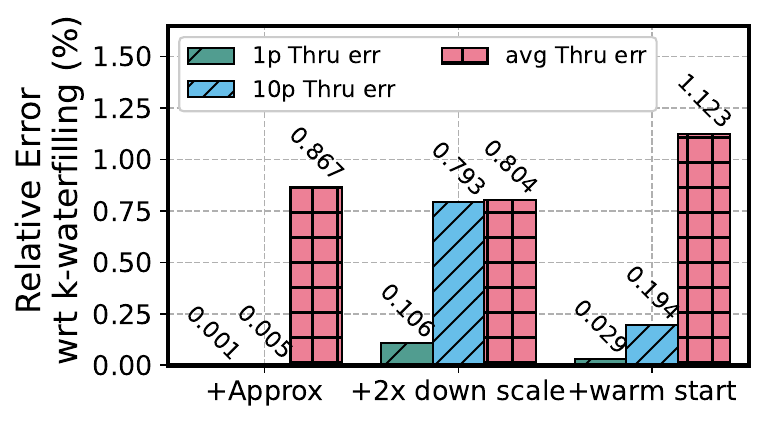} \label{fig:scalability:error}}
    \hfill
    \subfigure[Speed up of different scalability choices]{\includegraphics[width=0.33\linewidth]{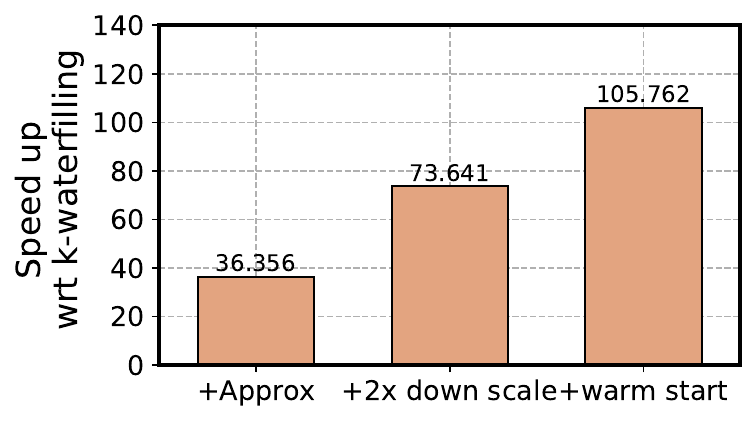} \label{fig:scalability:speedup}}
    \caption{\textbf{Scalability.} (a) \sysname scales almost linearly with the number of servers. (b) \cradd{Error introduced by each of our scaling techniques} (approx refers to the ultra-fast max-min fair algorithm). (c) Speed-up from each of our scaling components. \cradd{We compare with a version of \sysname that does not use the methods from~\secref{s:other-details}, instead relies on extended 1-waterfilling~\cite{Lavanya-sPerc} to compute max-min fair rates.}}
    \label{fig:scalability}
\end{figure*}

\parab{Scenario 3: Packet corruption at ToR.}
In this scenario, we consider (\tabref{tab:experiment-details}) failure patterns in which a ToR drops packets at either {\em high} \cadd{5\%} or {\em low} \cadd{0.005\%} rates. We also consider cases where both the ToR and a core link drop packets. CorrOpt and NetPilot do not support this failure as they can only account for scenarios where the network has redundant paths. The operator playbook makes a local decision on whether to mitigate the failure based on the severity of the packet loss, without considering the overall network condition. When the second failure occurs above the ToR, it reduces the capacity in the network core, causing the operator's approach to incur higher performance penalties.

Both playbook-based approaches suffer from performance penalties at least 2$\times$ higher than \sysname under PriorityFCT (\figref{fig:scenario-tor-drop}). \sysname has
a worst-case FCT penalty of 28.9\% while the best operator approach has a worst-case FCT penalty of 57\%. \sysname is again the only approach that achieves low penalty across all three metrics for both comparators.

\subsection{Other Results}
\label{s:eval-other}



\noindent{\bf Scalability.} In \figref{fig:scalability:run-time}, we show the time \sysname needs to find the best mitigation. \sysname's runtime scales linearly with the number of servers. Even on large-scale Clos topologies with 16K servers, \sysname needs less than 5 minutes, a fraction of the mitigation times operators report today~\cite{GCP-Incidents,Gao-Scouts}.



\sysname employs several approximation techniques to scale~\secref{s:other-details}. In~\figref{fig:scalability}, we quantify the error and speed-up of these methods compared to a version of \sysname without these approximations. Each technique in \secref{s:other-details} contributes significantly to speedup: (a) the max-min fair algorithm improves run-time by $36.3\times$ and only introduces $\le 0.9\%$ error; (b) \cadd{downscaling} traffic by $2\times$ does not introduce additional error but produces $73.6\times$ speedup! (c) adding warm-start and reducing the number of epochs results in $105.7\times$ speedup and $\le 1.2\%$ error. Future work can further speed this up by increasing the scale factor or reducing the number of epochs.

\parab{Sensitivity analysis.} We evaluate {\sysname}'s sensitivity to various inputs, including the drop rate estimates and the flow-arrival rates (see~\secref{sec::sensitivity} for details). 

As the inputs to the system vary, there are a few inflection points where \sysname is sensitive. These points are where \sysname can make mistakes and pick sub-optimal mitigations if the inputs are noisy. However, we find the difference between the impact of the mitigations is small around these inputs. Outside of these areas, the choice of better mitigation is clear. Thus, \sysname can tolerate large errors in the input distribution: the difference between mitigations is either large enough around that point so the choice of mitigation is clear or it is small enough where a mistake is not too costly.

\sysname can also pick the best mitigation under different congestion control protocols (see~\figref{fig:sensitivity:CC}). For this, we compare two protocols with different behaviors under loss: (1) Cubic~\cite{cubic}, which drastically reduces its sending rate under packet loss, and (2) BBR~\cite{bbr}, which does not. \sysname picks the best mitigation irrespective of which protocol we use. However. its approximations of the $1$p throughput distribution are more accurate when the mix of protocols is known (it can explicitly account for their differences in handling loss).  

\parab{Simulation validation.}
To show \sysname's effectiveness at larger scales with realistic link speeds and latencies, we conducted a simulation using \cadd{NS3}~\cite{henderson2008network} on a 128-server topology with 20~Gbps 100 $\mu$s links. \cadd{NS3} alone takes over a day to complete one simulation run (one sample). We invested significant effort in parallelizing \cadd{NS3} using MPI and reduced the time to run one sample to 6 hours. On this topology, we induce a failure in which two links drop packets (one ToR-T1 at 0.005\% and one T1-T2 at 0.5\%). This scenario shows the complex effect of different packet drops at different levels of datacenters and the trade-off between causing congestion by disabling the links versus incurring packet drops by taking no action. Apart from the DCTCP distribution, we also simulated FbHadoop~\cite{hadoop-workload} that has more short flows. These experiments required over 2100 hours to complete.

\begin{figure}[t]
  \centering
  \subfigure[DCTCP Traffic Distribution]{\includegraphics[width=1\linewidth]{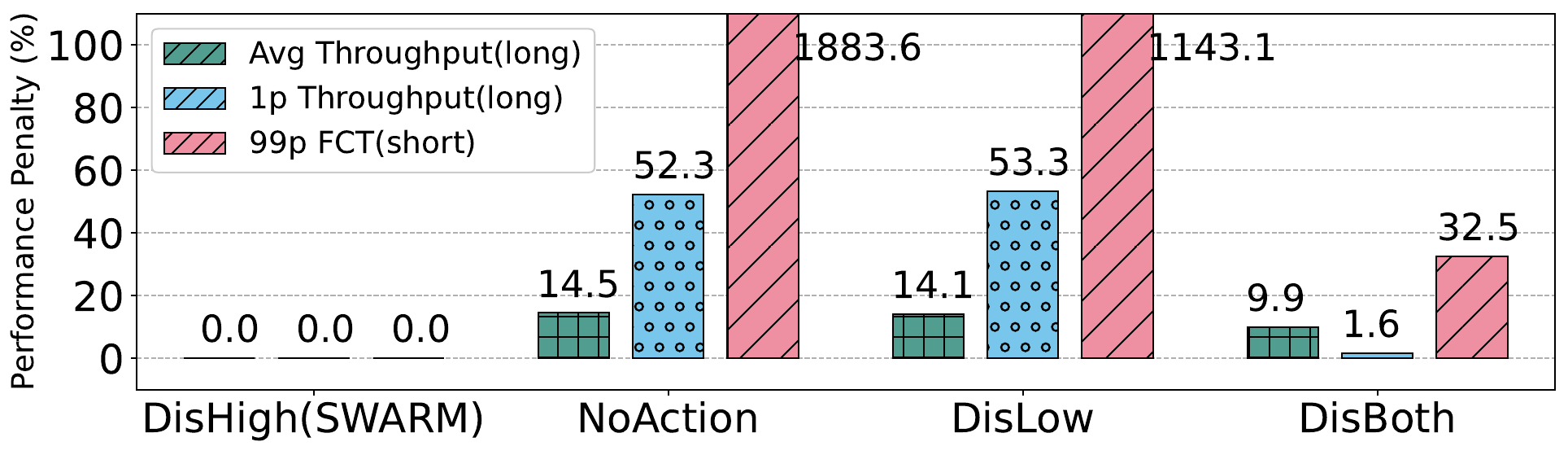}}
\\ 
\subfigure[FbHadoop Traffic Distribution]{\includegraphics[width=1\linewidth]{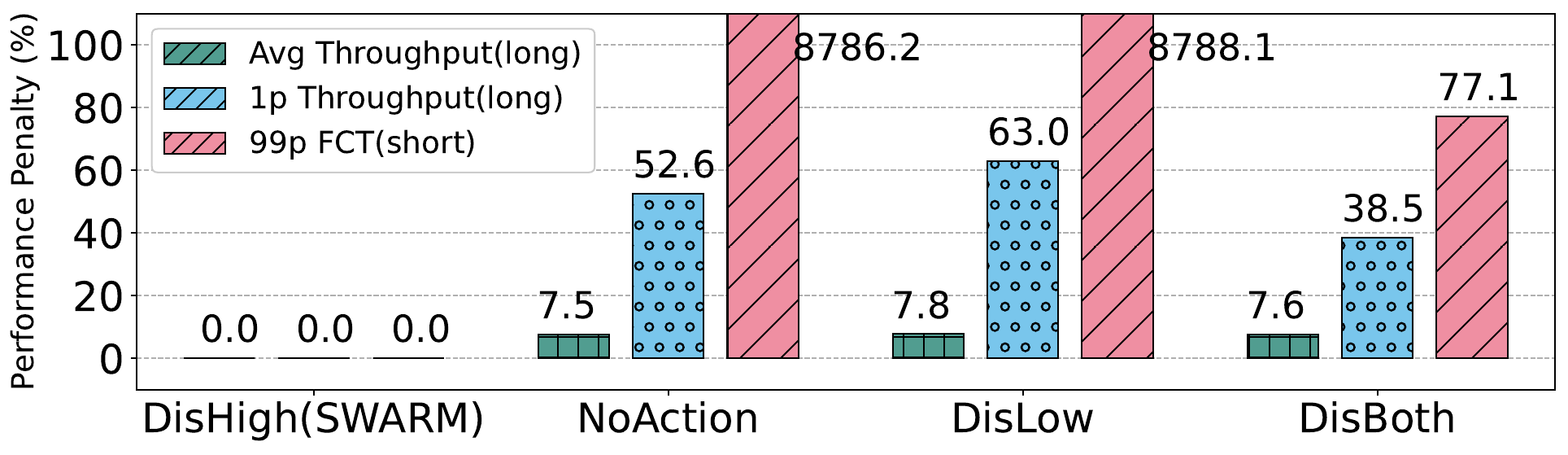}}
  \caption{\textbf{NS3 evaluation with different flow size distributions.} \cadd{(Dis=Disable, High=High drop link, Low=Low drop link)}\label{fig:ns3_sims}}
\end{figure}

\figref{fig:ns3_sims} shows the performance penalties of different actions. \sysname is able to identify if the congestion introduced by disabling the link would impact the flows more than the packet drop or vice-versa. In contrast, prior work (NetPilot, CorrOpt, and Operator) ignores the impact of traffic and failure characteristics on the mitigation. \sysname finds the best mitigation (only disabling the high drop rate link). In contrast, the baselines either disable both links or keep both links and incur 32\%~--~78\% penalty on 99p FCT. \cadd{Note that the performance penalty of not taking any action is the same as only disabling the link with a low packet drop since they both keep the link with a high drop rate in the network. This link ends up dropping many packets and determines the tail FCT.}

\parab{Testbed validation.}
To demonstrate \sysname makes high-quality decisions even on a physical testbed, we induce a failure pattern from Scenario 1 in which a ToR~--~T1 link randomly drops 6.25\% ($\frac{1}{16}$) of the packets, while a link from a different T1 to a T2 also drops packets at 0.39\% ($\frac{1}{256}$). 

\begin{figure}
    \centering
    \subfigure[Priority FCT]{\includegraphics[width=0.96\linewidth]{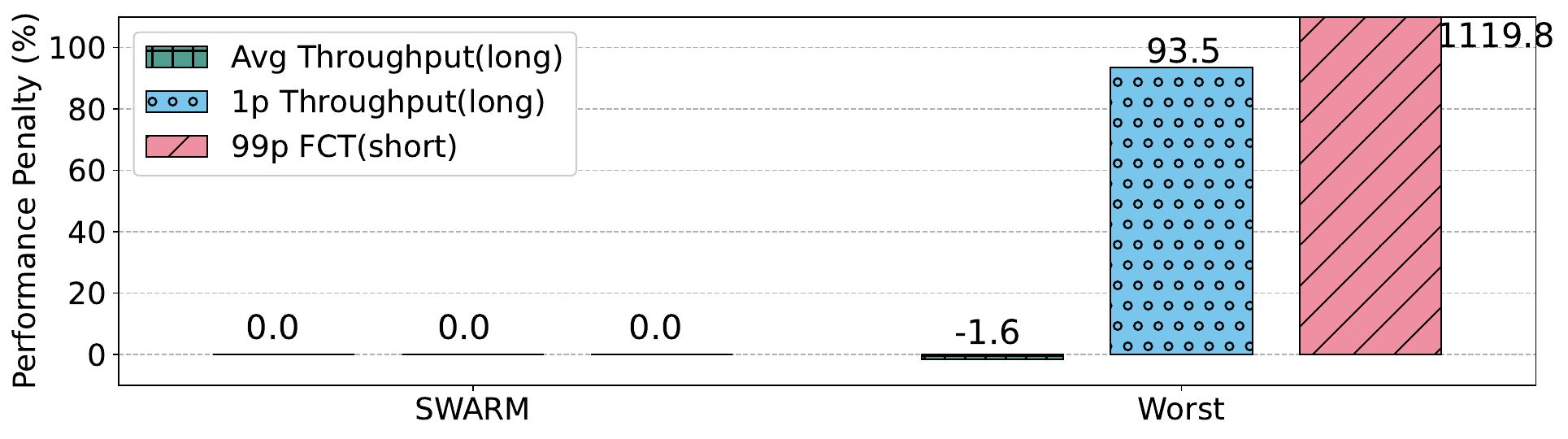}}
    \\ 
    \subfigure[Priority AvgT]{\includegraphics[width=0.96\linewidth]{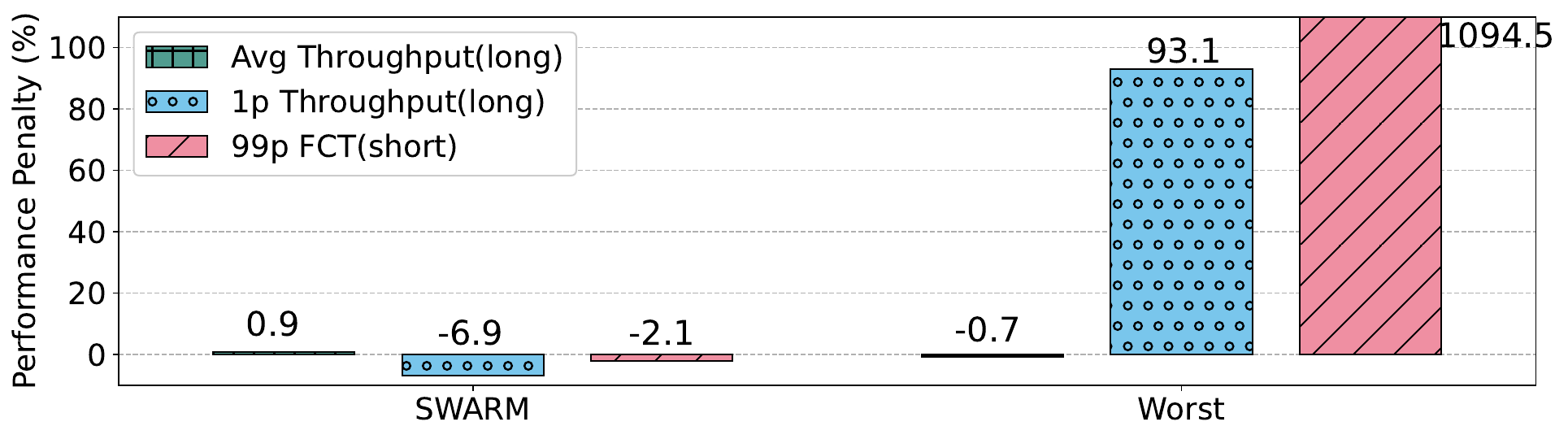}}
    \vspace{-1mm}
    \caption{\textbf{Physical testbed validation with two links dropping packets.} \cradd{We consider four mitigation strategies (disable or take no action). Optimal mitigation has zero penalty.}}
    \label{fig:testbed_scenario}
  \end{figure}

\sysname picks an optimal mitigation for Priority-FCT and a mitigation with less than 1\% penalty for PriorityAvgT (\figref{fig:testbed_scenario}). In contrast, the FCT performance penalty for choosing the worst action under PriorityFCT is over 1000\%. While the average throughput performance penalty is low across all mitigations in this incident, \sysname picks an action with a low penalty across all three metrics under the PriorityAvgT comparator. It avoids the 93\% penalty in 1p throughput and 1095\% penalty in 99p FCT for the worst action.


\parab{Impact of the comparator.} \cadd{We also observe that \sysname adjusts its decision based on the comparator. Across all the single link failures from Scenario 1 (\tabref{tab:experiment-details}), \sysname takes no action more often under the PriorityAvgT comparator than the PriorityFCT. This is because average throughput is sensitive to the remaining capacity in the network, so \sysname is more likely to keep the lossy link up.}

\parab{Justifying design choices.}
We also conducted several experiments to justify the special treatment of loss-limited flows, estimating distributions of FCT and throughput, using multiple epochs to capture flow dynamics, using distributional measures to determine the best mitigation, and the importance of accounting for queueing delay (see~\secref{sec::appendix_ablation}).



%% file: discussion.tex
\section{Discussion and Limitations}
\label{sec::discussion}

\sysname is a first step towards CLP-aware incident mitigation. It has limitations that future work can address.


\parab{Support for loss-less transport.} Cloud providers increasingly use RDMA~\cite{gao2021cloud,singhvi20201rma} for internal traffic. We can modify \sysname's \impact to (a) detect and account for pauses and (b) model loss-recovery approaches when losses happen.

\parab{Support for other routing protocols.} \sysname applies to any datacenter that relies on ECMP or WCMP for routing. These are commonly used in Clos topologies. Our experiments show the effectiveness of \sysname on different variants of Clos. Future work can extend this to topologies that use traffic engineering (\eg direct-connect~\cite{teh2020couder,jupiter-evolving})

\parab{Mitigations with transient effects.} Some actions introduce transient risk. For example, a switch may drop packets during reboot, causing a non-steady-state impact on long flows.

\parab{Approximate failure localization.} \sysname waits for operators or automation to localize the failure. It can instead use a spatial failure distribution which is available much sooner. This lowers the mean time to repair. \sysname relies on the correct input of the \textit{type} of failure. 

\parab{Impact on wide area network traffic.} WAN traffic is a small fraction of the datacenter traffic~\cite{firestone2018azure,benson2010network}, so we ignore its impact. When WAN traffic increases, we would need to account for larger RTT.

\parab{Other extensions.} Future work can extend \sysname to other metrics such as jitter and failures of software components such as load balancers. It can also account for estimated repair time in ranking mitigations, which can be challenging as incidents with vastly different repair times often have similar symptoms~\cite{Govindan-EvolveOrDie}. Future work can also explore the benefits of modeling more than two classes (short and long) of flows.

%% file: relatedwork.tex
\section{Related Work}
\label{seC:related_work}
\parab{Impact or risk estimation in networking.} No prior work considers CLP-aware failure mitigation but several estimate the impact or future risk of management operations. Janus~\cite{Alipourfard_Janus} focuses on risk estimation for datacenter management operations. RSS~\cite{Yiting_RSS} estimates the risk for backbone management. TEAVAR~\cite{Bogle_TEAVAR} and~\cite{Risk-Optical-Backbone, OSPF-Risk} route traffic in production wide area networks to minimize long-term impact.~\cite{Mitra-Stochastic-TE-Risk} models the risk of demand uncertainty and its impact on revenue in a WAN. Other approaches minimize the impact of failures without explicitly accounting for risk~\cite{FFC-HARRY, Arrow-Manya}.

NetPilot~\cite{Wu_NetPilot} and CorrOpt~\cite{Zhuo_CorrOpt} are closest to \sysname.  These can fail to produce mitigations with minimal CLP impact (see~\secref{sec:motivation}) because they (a) ignore the failure pattern, (b) do not account for traffic changes, (c) use proxy metrics (\eg maximum utilization) that only loosely correlate with CLPs, and (d) estimate impact on a healthy network, which limits the set of mitigations they support.

\parab{Computing fair share.} 
Prior work assumes flows follow max-min fairness, which matches the objective of TCP~\cite{Ben-StatisticalBWSharing}, and either employ optimization~\cite{Singla-Jellyfish, Namyar-TUB, SWAN, danna-practical-max-min,max-min}, use simulations~\cite{parsimon}, or iterative algorithms~\cite{Lavanya-sPerc, Jain-B4, Bottleneck-Ros-Giralt,danna-upward,ros2001theory,max-min} to find the fair share. These approaches are limited; they  assume flow rates are only limited by the network capacity and ignore packet drops. 

\section{Conclusion}

Failure mitigation is crucial in running datacenters at scale, but operators often find it hard to choose the right action. \sysname ranks mitigations based on their impact on well-known CLP metrics. Our evaluation shows approximating these mertics is possible at scale and results in mitigations with orders of magnitude lower performance penalty.

%% file: appendix.tex
\appendix
\renewcommand\thefigure{A.\arabic{figure}}
\renewcommand\thesection{\Alph{section}}
\renewcommand{\thealgocf}{A.\arabic{algocf}}
\renewcommand\thetable{A.\arabic{table}}
\setcounter{section}{0}
\setcounter{algocf}{0}
\setcounter{figure}{0}
\setcounter{table}{0}

\section{Algorithm Details}

\subsection{\impact Function}
\label{appendix:impactfunction}

\algoref{Alg:risk-main-function} shows the \impact in detail. It first uses the DKW inequality~\cite{DKWInequality} to compute the necessary number of samples to achieve a certain confidence level. It then adjusts the topology and the traffic based on the mitigation $\mitigationAction$. Finally, it divides the resulting traffic into short and long flows, and computes their CLPs separately on multiple routing samples (\secref{sec:swarm:internals}).

\RestyleAlgo{ruled}
\begin{algorithm}[h]
\setstretch{1.25}
\DontPrintSemicolon
\LinesNumbered
\caption{{\impact} Function.}
\label{Alg:risk-main-function}
\KwIn{$\traffic$. traffic matrix (src, dest, start time, size).}
\KwIn{$\topology$. network state (location, type, failures).} 
\KwIn{$\mitigationAction$. mitigation.}
\KwIn{$\confidenceLevel$. confidence threshold}
\KwOut{$\impactFlows_{l}$ = $\{\impactFlows_{l}^{ \mathcal{R}}\}$. $\impactFlows_{l}^{, \mathcal{R}}$ is the distribution of impact over all the long flows for routing sample $\mathcal{R}$.}
\KwOut{$\impactFlows_{s}$ = $\{\impactFlows_{s}^{ \mathcal{R}}\}$. $\impactFlows_{s}^{\mathcal{R}}$ is the distribution of impact over all the short flows for routing sample $\mathcal{R}$.}
$N \gets \text{num\_samples}(\confidenceLevel)$ \;
$\actionTopology, \actionTraffic \gets \text{apply\_mitigation}(\topology, \mitigationAction, \traffic)$ \;
$\shortflowtraffic, \longflowtraffic \gets \text{split\_traffic}(\actionTraffic)$\; \nllabel{3rd:line:split-long-short}
    \For{$n \in 1 \dots N$}{
        $\mathcal{R}_{n} \gets \text{get\_routing\_sample}(\actionTopology)$ \; \nllabel{7th:line:route-sampling}
        $\impactFlows_{l}^{\mathcal{R}_n} \gets \text{impact\_long\_flows}(\topology_{a}, \longflowtraffic, \mathcal{R}_{n})$ \; \nllabel{6th:line:long-flow}
        $\impactFlows_{s}^{\mathcal{R}_n} \gets \text{impact\_short\_flows}(\topology_{a}, \shortflowtraffic, \mathcal{R}_{n})$ \; \nllabel{7th:line:short-flow}
    
}
\Return{$\impactFlows_{l}$, $\impactFlows_{s}$}
\end{algorithm}

\subsection{Demand-Aware Max-Min Fair}
\label{sec:waterfilling}

\impact computes the impact of a given mitigation $\mitigationAction$ on the throughput of long flows in two steps. (1) it estimates the drop-limited throughput for each flow (\lineref{1:Line:get-drop-tput} in \algoref{Alg:Fair-Share-Throughput}), assuming there is no congestion and the packet drop enforces the maximum possible rate. (2) \impact computes the max-min fair rate of each flow and enforces these drop-limited throughputs as an upper limit on the rate (demand) of each flow (\lineref{2:Line:Max-Min-Fair}). This ensures that a flow does not receive more than its drop-limited throughput unless it is limited by its fair share when congestion is more severe than the drop rate.

Algorithms to compute max-min fairness~\cite{Lavanya-sPerc} assume demands are unbounded and are only limited by the network capacities. In \sysname, we develop a demand-aware extension of these algorithms that can enforce limits on each flow's rate (\algoref{Alg:Capacity-Limited-Link}). To achieve this, we augment the topology and add one virtual edge per flow with capacity set at the drop-limited rate. Then, we use existing algorithms to compute the network-wide max-min fair rates on this augmented topology. We can use the same method to enforce congestion control rate limits in the first few epochs of each flow in cases where the protocol's congestion window limits the flow's rate. 


We can use any of the variants of the $k$-waterfilling algorithm~\cite{Lavanya-sPerc} in the network-wide-max-min-fair function in \lineref{Line6:demand-aware:max-min} (but some can result in O(|E| + |F|) number of iterations where |E| is the number of edges and |F| the number of flows). Instead, we use a faster variant of max-min fairness~\cite{max-min} that speeds up k-waterfilling by 30$\times$ with minor degradation in the quality of the estimated rates (\figref{fig:scalability:speedup}, \figref{fig:scalability:error}).

{\small
\RestyleAlgo{ruled}
\begin{algorithm}[t]
\setstretch{1.25}
\DontPrintSemicolon
\LinesNumbered
\caption{Compute throughput.}
\label{Alg:Fair-Share-Throughput} 
\KwIn{$\topology$. current network state.}
\KwIn{$\traffic_{1\times n}$. source destination pairs (long flows).}
\KwIn{$\mathcal{R}_{1\times n}$. the (sampled) routing for each flow.}
\KwOut{$F$. max-min fair rate}
$\throughput \gets \text{get\_drop\_limited\_throughput}(\topology, \traffic, \mathcal{R})$ \; \nllabel{1:Line:get-drop-tput}

$F \gets \text{demand\_aware\_max\_min\_fair}(\topology, \traffic, \theta, \mathcal{R})$\; \nllabel{2:Line:Max-Min-Fair}
\Return{F}
\end{algorithm}
}

\RestyleAlgo{ruled}
\begin{algorithm}[t]
\setstretch{1.25}
\DontPrintSemicolon
\LinesNumbered
\caption{Demand-Aware Max-Min Fair.}
\label{Alg:Capacity-Limited-Link} 
\KwIn{$\topology$. current network state.}
\KwIn{$\traffic_{1\times n}$. source destination pairs (long flows).}
\KwIn{$\mathcal{R}_{1\times n}$. the (sampled) routing for each flow.}
\KwIn{$\throughput_{1 \times n}$. the drop-limited rate for each flow.}
\KwOut{$\hat{\throughput}_{1 \times n}$. max-min fair rates after applying congestion.}
\For{$\forall f \in \traffic$}{
    \tcc{add one virtual link per flow}
    $e \gets G.\text{add\_virtual\_edge}()$ \;
    \tcc{set capacity to drop-limited rate}
    $e.\text{capacity} = \throughput_{f}$ \;
    \tcc{add to the route of flow $f$}
    $\mathcal{R}_{f}.\text{add}(e)$ \;
}
$\hat{\throughput} \gets \text{network\_wide\_max\_min\_fairness}(\topology, \traffic, \mathcal{R})$ \; \nllabel{Line6:demand-aware:max-min}
\Return{$\hat{\throughput}$}
\end{algorithm}

\section{Details of Offline Measurements}
\label{appendix:offline-measurement}
\label{appendix:queueing}

\begin{figure*}
    \centering
    \includegraphics{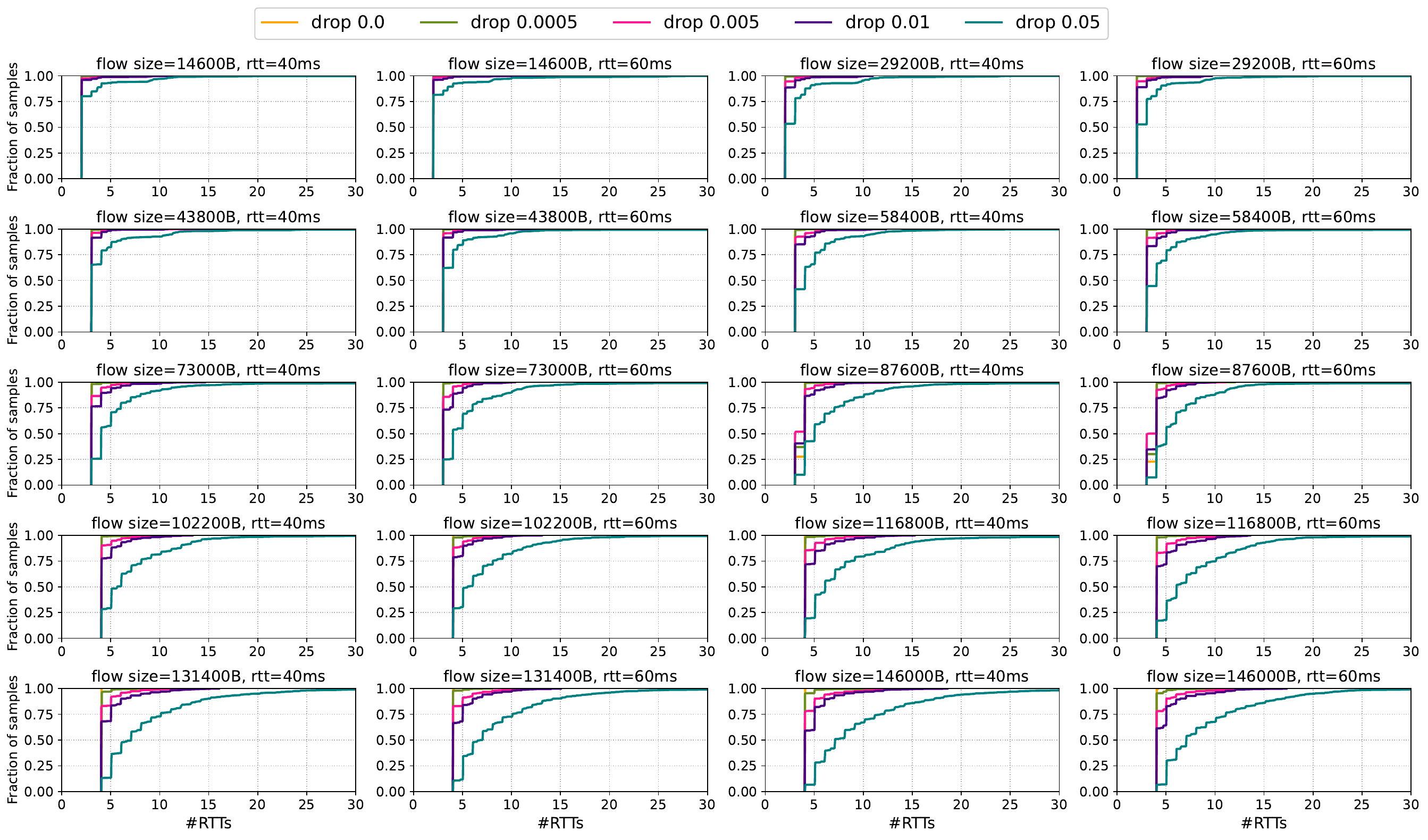}
    \caption{\textbf{An example of distribution measured for short flow's FCT \cadd{using Mininet}.}}
    \label{fig:example-distribution:rtt}
\end{figure*}

We measure three empirically driven distributions offline and use them to estimate CLPs in \sysname: (1) the upper bound on the throughput of long flows in a lossy network, (2) the necessary number of RTTs to deliver short flows, and (3) the queueing delay. In this section, we describe how we gathered these distributions. We also show a sample in \figref{fig:example-distribution:rtt}.

\begin{figure}
    \centering
    \subfigure[Topology 1]{\includegraphics[width=0.8\linewidth]{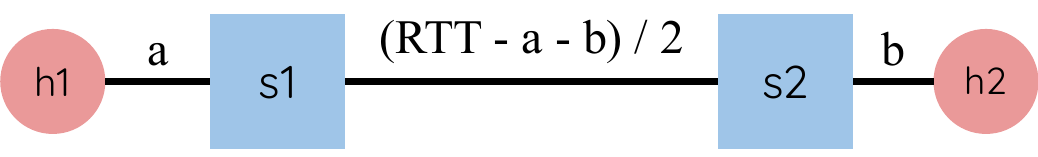}\label{fig:offline:topo1}}
    \subfigure[Topology 2]{\includegraphics[width=0.8\linewidth]{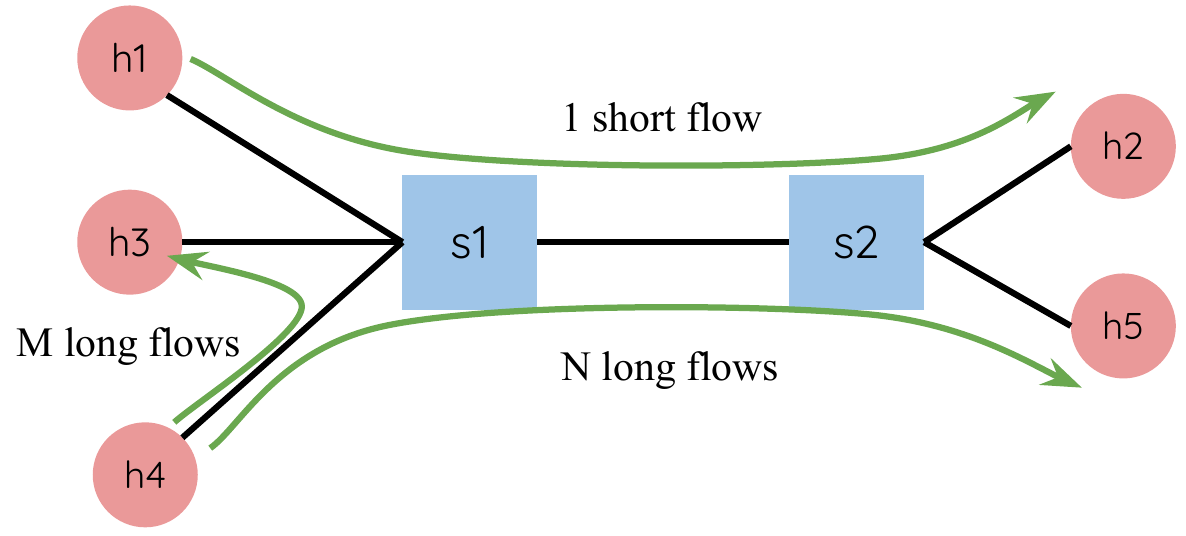}\label{fig:offline:topo2}}
    \caption{Topologies for offline measurements.}
    \label{fig:offline:topologies}
\end{figure}

\parab{Throughput of long flows in a lossy network.} We use \cadd{Topology 1 in \figref{fig:offline:topologies}} where h1 sends a long flow $f$ to h2. We use iperf3 to measure throughput under different network conditions (\eg by inducing packet drops in s1~--~s2 or changing the RTT by adjusting the delay on s1~--~s2). We ensure that link capacities are sufficiently high so that the drop rate always dictates the rate. We also repeat each experiment multiple times to create a statistically significant distribution~\cite{DKWInequality}.

\parab{Number of RTTs for short flows.} We send a short flow from h1 to h2 in Topology 1 (\figref{fig:offline:topologies}) under different settings (\eg flow size, slow start threshold, initial congestion window) and network conditions (\eg packet drop rate, RTT). We repeat each experiment multiple times, measure the FCT of the flow in each experiment, and construct an FCT distribution. We ignore the queueing delay since we only have one flow in this experiment and divide the FCT distribution by the two-way propagation delay to derive the necessary number of RTTs.

\parab{Queueing delay for short flows.}  Queueing delay is harder to measure as it depends on network load and the number of competing flows. We use Topology 2 in~\figref{fig:offline:topologies} to estimate the queueing delay distribution under various network conditions. We route M long flows from h4 to h3 and N additional long flows from h4 to h5. After these long flows reach the steady state, we route a flow small enough to finish within an RTT from h1 to h2 and measure its FCT. 

In this experiment, we can control the number of flows that compete on the s1~--~s2 link using N and the utilization of the s1~--~s2 link using M and N. The latter is because the h4~--~s1 is the main bottleneck in the network and determines the rate of the N long flows that go through s1~--~s2. We vary both M and N to model different network conditions and measure the FCT of the short flow. We subtract the two-way propagation delay from the FCT to derive the queueing delay.
Once again, we repeat these experiments to ensure statistical significance in the resulting distributions.

\begin{table*}[t]
\centering
{\footnotesize
\begin{tabularx}{\textwidth}{|m{0.5em}|c|X|c|} \hline
{} & {} & {\bf Details} & {\bf \#scenarios}\\ \hline \hline
{\parbox[t]{2mm}{\multirow{5}{*}{\rotatebox[origin=c]{90}{Scen. 1}}}} & {1 link failure} & {One T0~--~T1 and one T1~--T2, each with two levels of packet drop rate \cadd{$\sim$5\%} and \cadd{0.005\%}} & {4} \\\cline{2-4}
{} & {\multirow{4}{*}{2 link failures}} & {Four combination of pair of links (two T0~--~T1s in the same cluster connected to the same T0, two T0~--~T1s in the same cluster connected to different T0s \& T1s, one T0~--~T1 \& one T1~--~T2 connected to different T1s, and two T1~--~T2s connected to different T1s \& T2s), each with all combinations of the two packet drop levels and all the possible ordering of failures.} & {\multirow{4}{*}{32}}\\ \hline \hline

{\parbox[t]{2mm}{\multirow{3}{*}{\rotatebox[origin=c]{90}{Scen. 2}}}} & {no other link failure} & {one T1~--~T2 with capacity reduced to half} & {1} \\ \cline{2-4}
{} & {\multirow{2}{*}{1 other link failure}} & {one T1~--~T2 with capacity reduced to half, another T0~--T1 link with 3 levels of failure (two packet drop levels + completely down), and all possible ordering of failures.} & {\multirow{2}{*}{6}}
\\ \hline \hline

{\parbox[t]{2mm}{\multirow{4}{*}{\rotatebox[origin=c]{90}{Scen. 3}}}} & {no link failure} & {One T0 with two levels of packet drop rate \cadd{$\sim$5\%} and \cadd{0.005\%}} & {2} \\ \cline{2-4}
{} & {\multirow{3}{*}{1 link failure}} & {One T0 and One T0~--~T1 in the same cluster connected to a different T0, with all combinations of packet drop rates (ToR with two levels of packet drop rate and link with three levels of failure (two packet drop rates + completely down)), and all the possible ordering of failures.} & \multirow{3}{*}{12}
\\ \hline \hline
{} & {} & total number of evaluated scenarios & 57
\\ \hline
\end{tabularx}}
\caption{Mininet Experiment Details.}
\label{tab:experiment-details}
\end{table*}

\section{Additional Experiment Details}
\label{appendix:experiment-details}
In this section, we describe the details of how we evaluate \sysname using emulation, simulation, and testbed experiments on a set of representative and common incidents.

\subsection{Traffic Traces}
\label{a:traffic-trace}
We evaluate our approach using many traffic traces to ensure our results are reliable. Each trace includes several network flows, each of which has 4 pieces of information: source, destination, size, and start time. We ensure these traces are long enough to capture the impact of mitigation on a wide range of flows with various sizes. To prevent capturing the impact of an empty network, we discard the performance of flows that begin in an initial window of the trace. \cadd{Note that the duration of this window and the trace itself depend on the network bandwidth and delay in each setup.}

\parab{Flow sizes.} We follow recent works~\cite{hpcc,tpprof} and sample flow sizes from a well-known and widely used distribution from DCTCP~\cite{Alizadeh-DCTCP}.

\parab{Flow start time.} We generate flow start times using a Poisson distribution~\cite{BeyondFattrees,ExpandTime,teh2020couder,hpcc} with inter-arrival time derived from \azure to ensure the load on the network is reasonable (unless otherwise specified).

\parab{Source and destination.} We follow~\cite{hpcc} in picking the source and destinations.

\subsection{Failures Types}
\label{a:failure}

We evaluate our approach on different types of failures (see~\tabref{Table::failurelist}). These failures (\eg packet drop on multiple links or capacity drop due to fiber cut) are common in \azure and other cloud providers, as several studies confirm~\cite{Arzani-007,omnimon,Roy-NSDI17,Gill-UnderstandingFailures,Zhuo_CorrOpt,Wu_NetPilot}.

We use Mininet (a network emulator) as our primary evaluation framework for the reasons outlined below. We provide specific details on the failure cases in \tabref{tab:experiment-details}. These 57 cases represent a wide range of potential issues that could arise in a datacenter. For instance, we evaluate two scenarios per packet drop rate for single-link failure incidents, which cover all possible single-link failures that can occur in a data center. This is because (a) the target link is either between ToR and an aggregation or between an aggregation and a core switch~\cite{fattree}, and (b) Clos is a symmetric topology. Therefore, all ToR-aggregation and core-aggregation links are equivalent~\cite{Alipourfard_Janus}. This means it is enough to only look at two cases, which are representative of a wide range of possible incidents.

The same logic applies to other cases. For example, one of our two-link failure scenarios considers the case where two ToR-aggregation links in the same pod drop packets. This scenario represents any two ToR-aggregation links that may fail in any of the pods over the entire datacenter.

\subsection{Emulation, Simulation, and Testbed}
\label{a:setup}

\parab{Emulation Setup (Mininet).} Mininet uses a real,
production-grade TCP/IP stack from the Linux kernel. This allows us to evaluate our methods on production-grade implementations of congestion control algorithms such as BBR~\cite{bbr} or Cubic~\cite{cubic}. However, scaling Mininet is challenging~\cite{Bottleneck-Ros-Giralt}. We use the Clos topology from~\figref{fig:example_two_drops_motiv} with 8 servers, 4 ToRs, 4 aggregation switches (T1), and 4 spine switches (T2). Even with this setup, we required a server with 64 cores and \cadd{256~GB} memory to achieve the desired load level. We include results from {\it 4000 hours experiments (over 5 months)} to cover all 57 scenarios, all combinations of possible mitigations, and running each multiple times.

Emulating a network with gigabit-level bandwidth (40 Gbps) and microsecond-level propagation delay (50 $\mu$s) is impossible in Mininet. Therefore, we downscale the traffic and the link properties following~\cite{Shrink-Pan,Konstantinos_downscale_simulation}. This involves (1) reducing the link capacities and increasing the link delay to ensure the network bandwidth-delay product remains the same and (2) increasing the inter-arrival time between flows by the same factor (in our case, 120$\times$). Our testbed and simulation experiments compensate for this.

We tested our approach with both Cubic~\cite{cubic} and BBR~\cite{bbr} congestion control algorithms in Mininet (and DCTCP~\cite{Alizadeh-DCTCP} in our NS3 simulations). Our findings suggest that the choice of mitigation depends less on the congestion control, and a common abstraction that ensures these flows receive max-min fair rates is enough (TCP objective~\cite{Ben-StatisticalBWSharing}).

For our evaluation, we use traffic traces that last 500 seconds (around 9 minutes) to capture enough samples from different flow sizes. We only measure the impact on flows that start within [50, 150) seconds to avoid including the start-up phase and to ensure all the captured flows finish before the end of the traffic trace. We repeat each experiment for 30 different traffic traces (\secref{a:traffic-trace}).

\parab{Simulation Setup (NS3).} We use a Clos topology with 16 cores (T2), 32 aggregations (T1), 32 ToRs, and 128 servers. Each link has 20~Gbps bandwidth and 100~$\mu$s propagation delay. We use DCTCP~\cite{Alizadeh-DCTCP} as the congestion control algorithm.

We consider an example where two links drop packets (one at a high rate \cadd{$\sim$0.5\%} and the other one at a low rate \cadd{$\sim$0.005\%}). This scenario captures the impact of both the drop rates and the network load on the decision. We also find the packet drop rate impacts the scalability of the simulation. Thus, we had to slightly reduce it compared to our Mininet experiments. Even with this drop rate, it takes NS3 one day to compute the performance of one mitigation on a single traffic trace. We use MPI to speed up NS3 and reduce the runtime.

Each of our traffic traces is 10 s long, and we measure the impact on flows that start within [0.5, 1) s. We repeat each experiment for 30 different traffic traces.

\parab{Testbed Setup.} Our testbed has 32 servers connected using a Clos topology with six ToRs, four aggregation switches (T1), and two core switches (T2). Each link in our testbed has 10~Gbps bandwidth and 200~$\mu$s propagation delay. The topology is a Clos~\cite{fattree} where all T1 and T2 switches are connected to each other (different from Mininet and NS3 simulations).

Due to hardware limitations, we can only inject packet drops at the power two. We consider an example where we have two links dropping packets (one at a high rate $\frac{1}{16}$ and the other one at a low rate $\frac{1}{256}$).

\subsection{Other Details}
\label{a:exp-other-details}
\parab{\sysname}. We use 32 traffic traces and 1000 routing samples in \sysname (unless mentioned otherwise). For our baseline comparisons in Mininet, we set the duration of each traffic trace to 200~s, and as in Mininet, we compute the CLP metrics over all the flows that start within [50, 150) s. We also use 200 ms as the epoch size for \algoref{Alg:Impact-Long-Flows}. Note that an ideal epoch size should be in the order of the flow inter-arrival time (1~ms in Mininet). However, we find that \sysname can still find good mitigations even when we use a much larger epoch size.

\parab{Routing.} We assume either ECMP or WCMP routing~\cite{wcmp,VL2-Microsoft}, which are typical in data centers.

\section{Extended Evaluation}
\label{sec:extended_eval}
In this section, we present an extended evaluation of \sysname.

\subsection{Sensitivity to Inputs}
\label{sec::sensitivity}

\begin{figure}
    \centering
    \subfigure[Sensitivity to packet drop rate]{\includegraphics[width=0.9\linewidth]{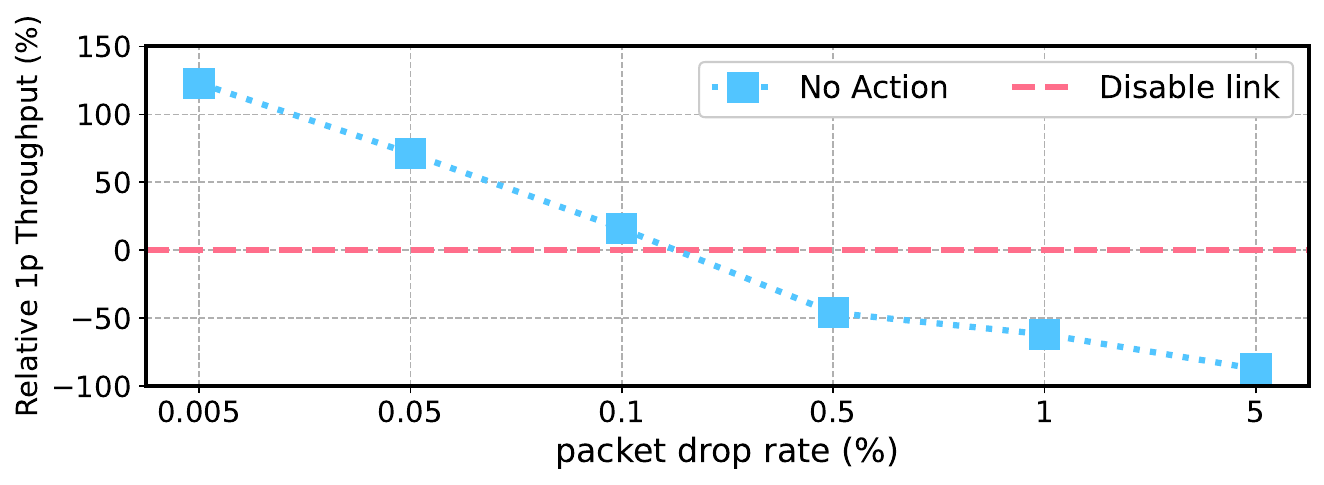} \label{fig:sensitivity:drop_rate}} 
    \\
    \subfigure[Sensitivity to flow arrival rate changes]{\includegraphics[width=0.9\linewidth]{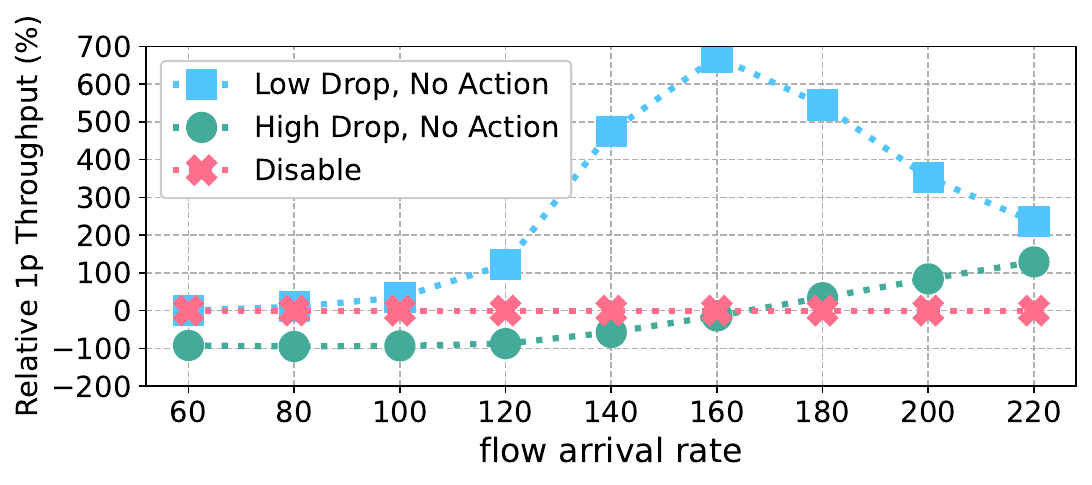} \label{fig:sensitivity:arrival_rate}}
    \caption{\textbf{Evaluating sensitivity of \sysname to errors in the input packet drop rate and flow arrival rate \cadd{using Mininet}.} All experiments are in a scenario where a T0~--~T1 drop packets.}
    \label{fig:sensitivity}
\end{figure}


\parab{Packet drop rates.} 
We compare the relative difference between the 1p throughput of taking no action and disabling a link in a failure scenario where a T0~--~T1 link drops packets at different rates (\figref{fig:sensitivity}). We find the choice of the right decision to be bi-modal with a wide room for error. It is better to take no action for all the drop rates below $\sim 0.1\%$ while the best action is to disable the link beyond that point. Also, the penalty close to this transition point ($0.1\%$ drop rate) is rather small. In other words, the error in the input packet loss rate has to be an order of magnitude for \sysname to make the wrong decision, an unlikely possibility in today's clouds.


\parab{Flow arrival rates.} We investigate how \sysname's decisions change under two failure severities and different flow arrival rates (\figref{fig:sensitivity}). We observe that the gap between the two decisions is significant outside of a few inflection points, which means \sysname will pick the best mitigation in most cases.

For instance, in the high drop rate scenario, disabling the link is better than taking no action for arrival rates ranging from 60 to 160 flows per second (fps). Taking no action is better once we pass 160 fps as bringing down the link causes network congestion. This means \sysname has a wide margin of error. At small and large arrival rates, the difference between the two actions is consistently large, which means \sysname would pick the right action. At medium arrival rates, the difference is very small, so whichever action \sysname picks results in almost the same impact.

\subsection{Sensitivity to Congestion Control}
\label{a:sensitivity-congestion}
\begin{figure}
    \centering
    \includegraphics[width=1.0\linewidth]{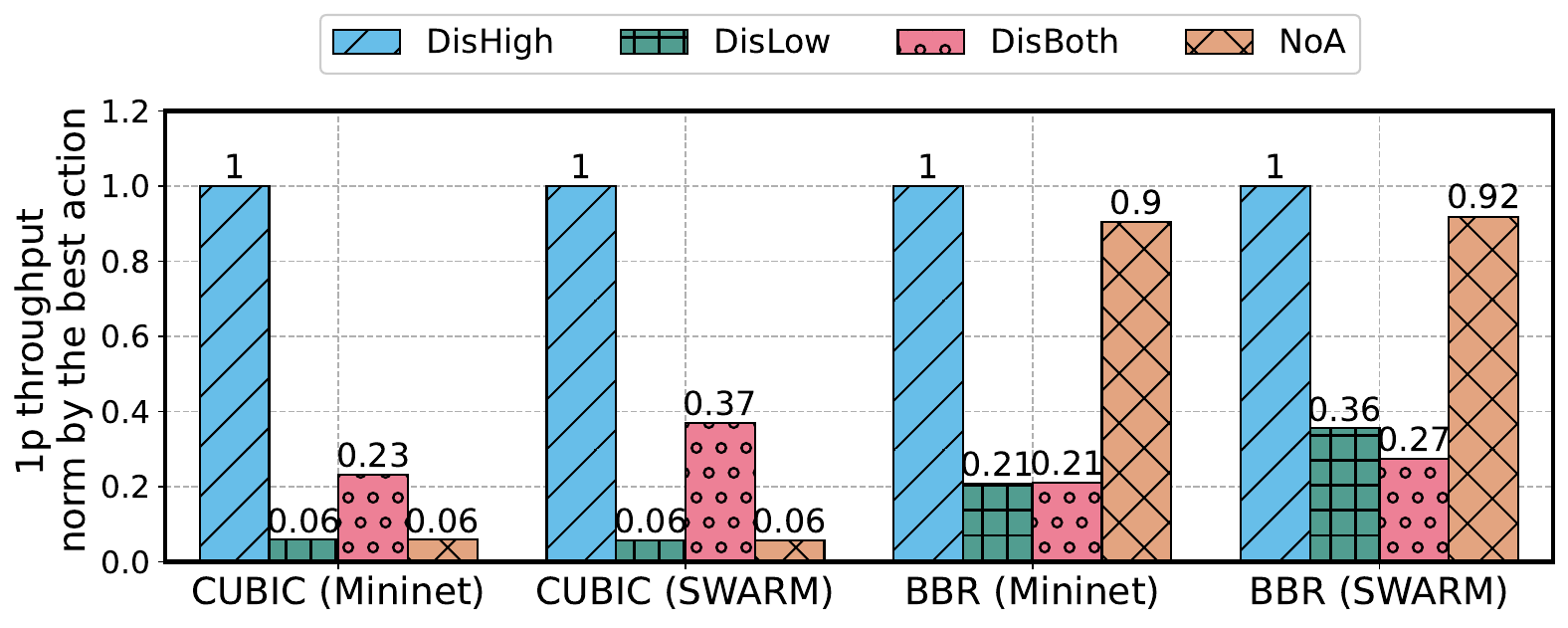}\label{fig:generalization:CC}
\caption{\textbf{Evaluation of \sysname on multiple congestion controls \cadd{in Mininet}.} In general, \sysname is able to correctly order the mitigation actions and approximately capture the relative difference between different mitigation actions on both congestion controls (BBR~\cite{bbr} and Cubic~\cite{cubic}). }
    \label{fig:sensitivity:CC}
\end{figure}

We conduct a limited experiment to evaluate whether \sysname is resilient to the choice of congestion control. We consider a scenario where a T0~--~T1 and a T1~--~T2 are dropping packets at low and high drop rates, respectively. We use two example congestion control protocols (CUBIC~\cite{cubic} and BBR~\cite{bbr}), which behave differently under loss. Cubic significantly reduces its rate under loss while BBR does not. 

We compare {\sysname}'s estimated 1p throughput with Mininet under four mitigations (\figref{fig:sensitivity:CC}). We find (1) \sysname can estimate throughput adequately enough to pick the best mitigation and (2) the choice of the best action remains independent of the congestion control protocol. {\sysname}'s estimates are more accurate when it operates on the same congestion control protocol. This means it is more robust when operators correctly input the likelihood of different protocols being used inside their datacenters.

\begin{figure}
    \centering
    \subfigure{\includegraphics[width=.9\linewidth]{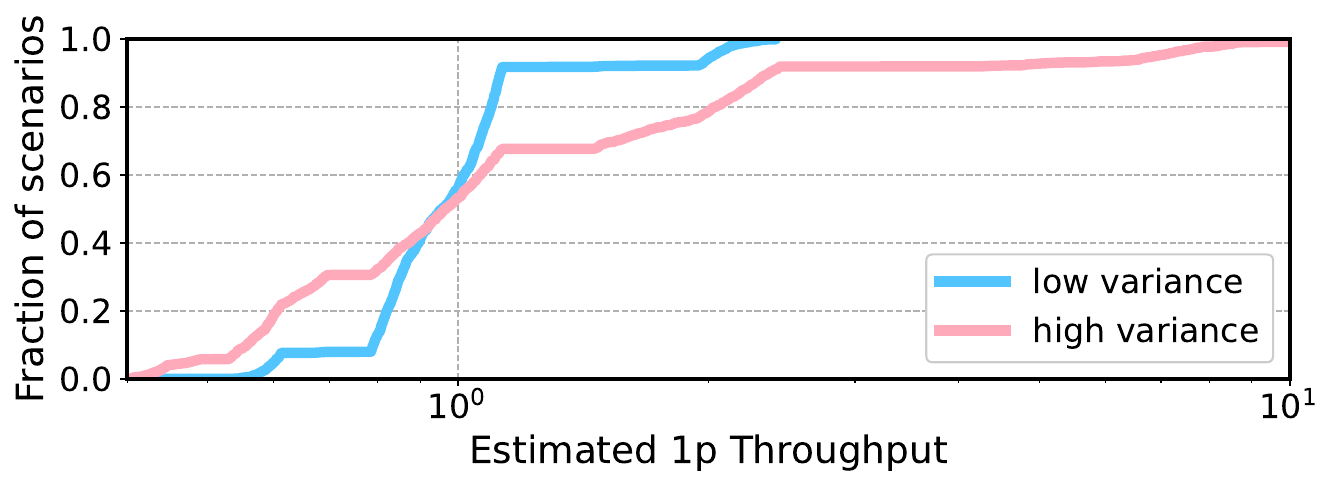}}
    \\ \vspace{-2mm}
    \subfigure{\includegraphics[width=.9\linewidth]{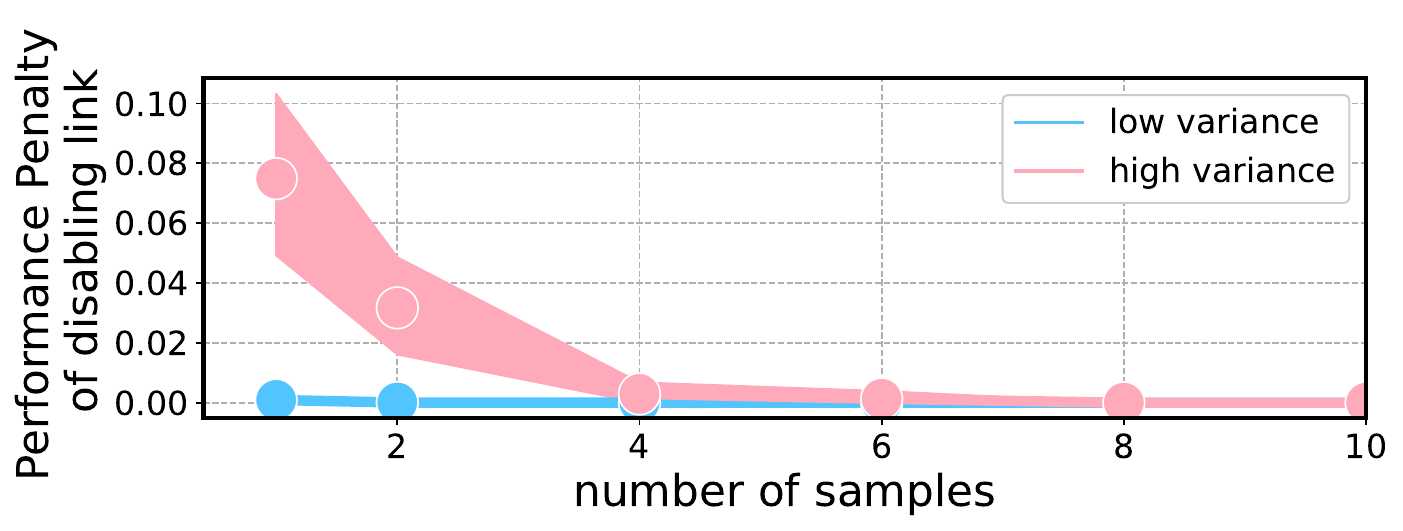}}
    \caption{Variance in the input flow arrival rate and its impact on the performance penalty of mitigation actions \cadd{in Mininet}.}
    \label{fig:variance-flow-arrival}
\end{figure}


\subsection{Assumptions and Design Choices}
\label{sec::appendix_ablation}
We also validate our assumptions and design choices.

\begin{figure}[t]
    \centering
    \subfigure[Drop-limited vs Capacity-limited]{\includegraphics[width=0.95\linewidth]{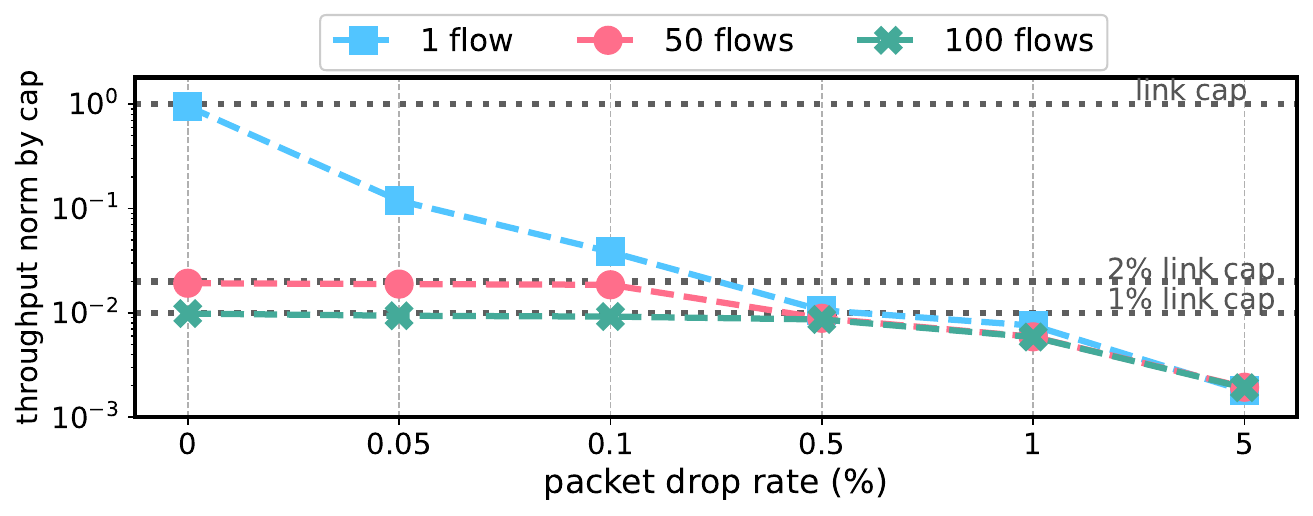}\label{fig:ablation:drop-capacity-limited}}
    \\
    \subfigure[Impact of different design choices on throughput estimation (E=Epoch, R=Routing Sample, T=Traffic Sample, S=Single, M=Multiple). For example, SE/SR/ST refers to using a single epoch, with a single routing sample, and a single traffic sample. We run each method with 10 different seeds.]{\includegraphics[width=0.95\linewidth]{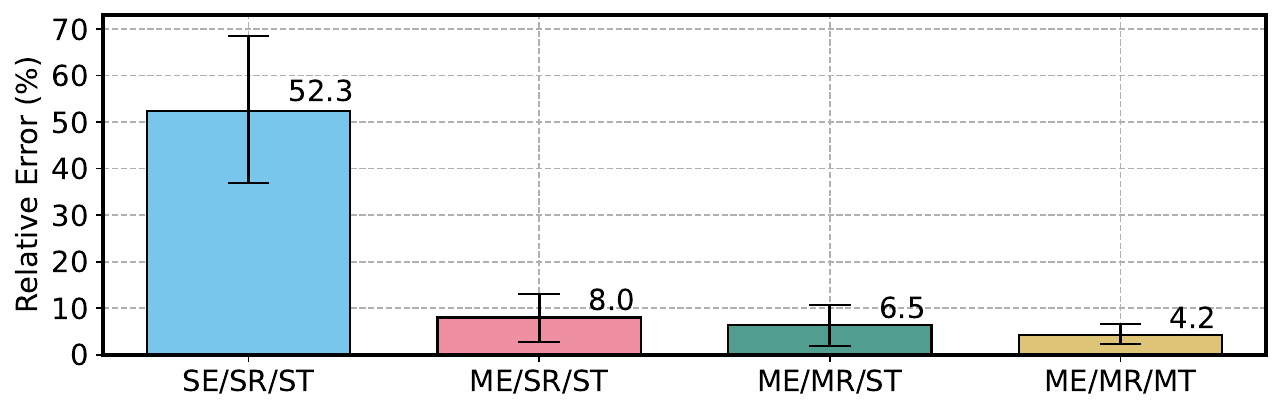} \label{fig:ablation:multi-epoch-multi-sample}
    }
    \\
    \subtable[How queueing delay influences the mitigation choice.]{
        \centering
        \begin{tabular}{|l|l|c|}
            \hline
            Approach & Best Action & FCT Penalty\\ \hline \hline
            Ignore Queueing &  Disable C0~--~B0 & 48 \% \\ \hline
            Model Queueing &  Bring back C0~--~B0 & 0 \% \\ \hline
        \end{tabular}
        \label{tab:ablation:queue-delay}
    }
    \caption{\textbf{Validationg assumptions and design choices \cadd{using Mininet}.} (a) shows flows are capacity- or loss-limited. A flow is loss-limited when its rate drops below its fair share of the link capacity (marked as dashed lines) (b) shows the impact of our design choices and the relative error with respect to Mininet. (c) the importance of accounting for queuing delay.}
    \label{fig:ablation-study}
\end{figure}

\parab{Capturing the impact of drop-limited flows.} We account for drop-limited flows by computing an upper bound on their rate and then enforce these bounds during max-min fair rate computations. We evaluate this assumption in an experiment where a single link delivers a varying number of flows and drops packets at different rates (\figref{fig:ablation-study}). We observe our observation is valid: each flow takes the minimum of its fair share rate and its drop-limited throughput.

We next evaluate our design choices for estimating the distribution of the flow completion time and throughput (\figref{fig:ablation-study}). Specifically, we show the quality of {\sysname}'s approximations compared to Mininet measurements. We use a scenario where two links drop packets at different rates, and the mitigation is to disable the link with the higher drop rate.

\parab{Single Epoch vs. Multiple Epochs.} In \sysname, we use epochs to capture the shift in network bottlenecks as flows arrive or depart over time. We observe that not capturing these dynamics leads to more than 50\% estimation error on average (this is equivalent to running \sysname with only one epoch). Instead, \sysname uses multiple epochs to model flow arrival/departure and avoids this error.


\parab{Distributions vs deterministic estimates.} In \sysname, we pick the best mitigation based on CLP distributions. This method increases confidence in the final decision and has less error compared to relying on a single CLP sample. \sysname can efficiently compute these distributions in parallel.


\parab{Queueing Delay.} We show the importance of accounting for queuing delay using an example scenario where the link C0~--~B0 in the \figref{fig:example_two_drops_motiv} topology drops packets at a high drop rate and the operator prioritizes 99p FCT of short flows. In this case, the best action is to disable the link. After we apply the mitigation, the link C0~--~B1 starts to drop packets at a high drop rate. Disabling C0~--~B1 is not feasible anymore as it partitions the network, so we have to either take no action or bring back C0~--~B0. If we ignore the queuing delay, both of these options have the same approximate 99p FCT. However, bringing back C0~--~B0 is a better mitigation since it increases the path diversity and reduces the queuing delay (\tabref{fig:ablation-study}).

\subsection{Results for Other Comparators}
\label{appendix:other-comparators}

\parab{Priority1pT.} This comparator minimizes the 1p (\cadd{1st} percentile) throughput. It uses two tiebreakers in the following order: average throughput and 99p FCT.

\begin{figure*}[t]
  \centering
  \subfigure{\includegraphics[width=0.8\linewidth]{figures/Type1_incident_nsdi/legend_incident1-eps-converted-to.pdf}}
    \addtocounter{subfigure}{-1} \vspace{-4mm}\\
    \subfigure[Scenario 1]{\includegraphics[width=0.95\linewidth]{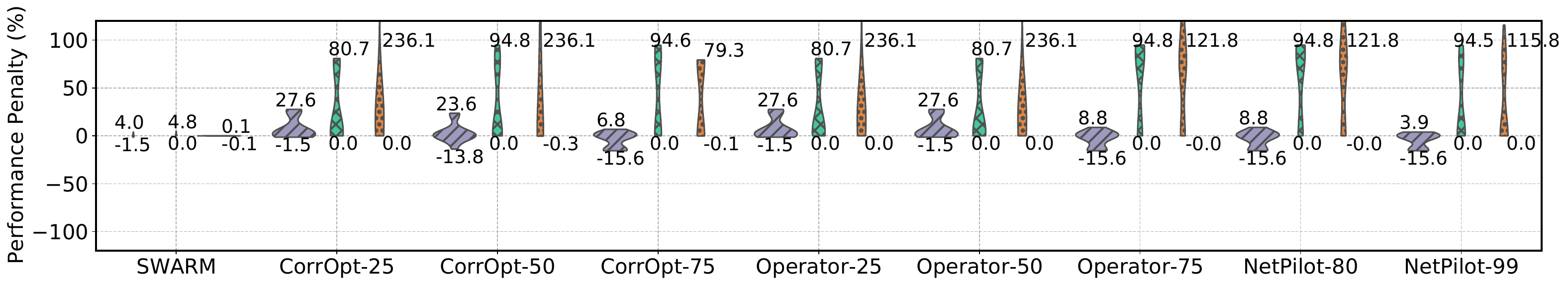}\label{fig:scenario_1_results:priority1pT}} \\ 
  \subfigure[Scenario 2]{\label{scenario_2_results:priority1pT}\includegraphics[width=0.45\textwidth]{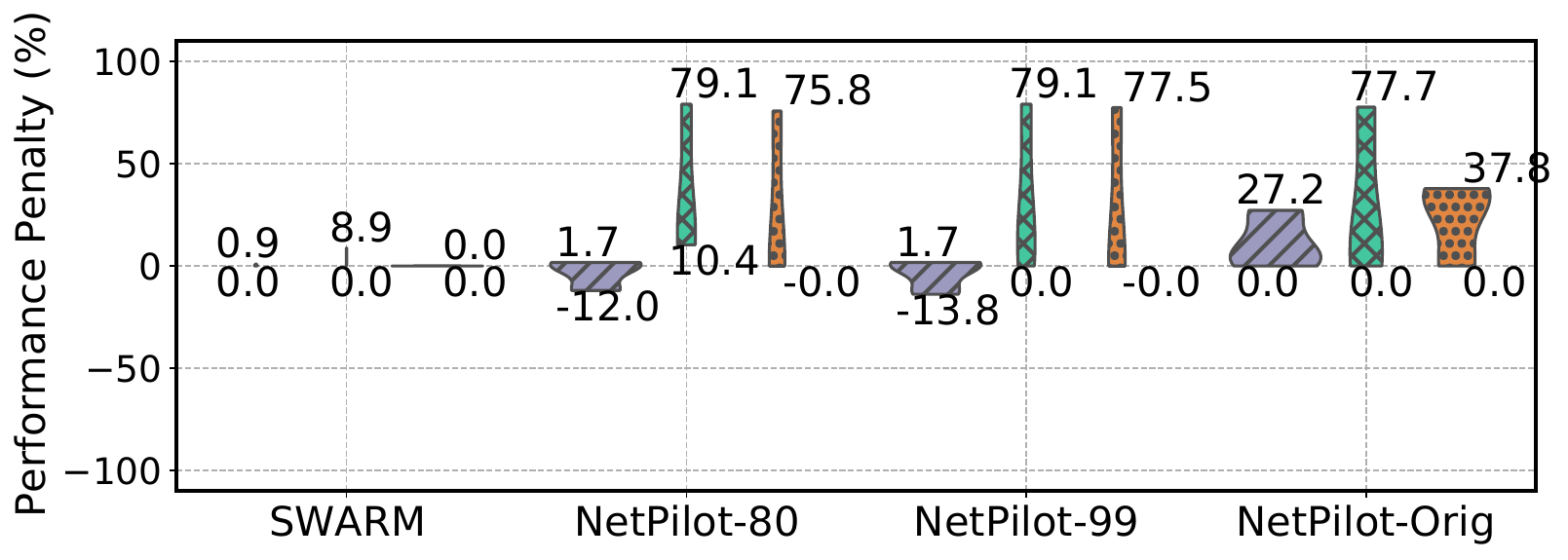}}
  \subfigure[Scenario 3]{\label{scenario_3_results:priority1pT}\includegraphics[width=0.45\textwidth]{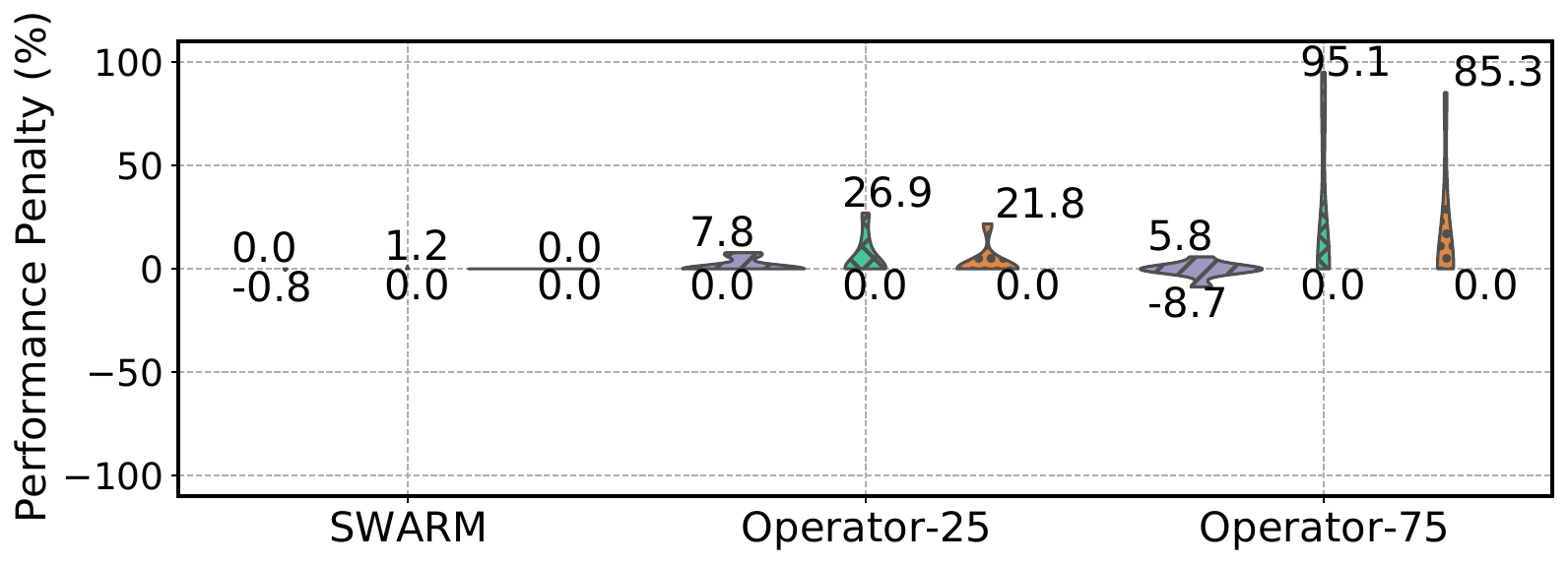}} \\
  \caption{\textbf{\sysname outperforms other baselines under Priority1pT comparator \cadd{in Mininet}.} It is the only method with a low penalty across all the metrics and all the scenarios.}
  \label{fig:scen-priority-1pT}
\end{figure*}

\begin{figure*}[h]
  \centering
  \subfigure{\includegraphics[width=0.8\linewidth]{figures/Type1_incident_nsdi/legend_incident1-eps-converted-to.pdf}}
    \addtocounter{subfigure}{-1} \vspace{-4mm}\\
  \subfigure[Scenario 1]{\includegraphics[width=0.95\linewidth]{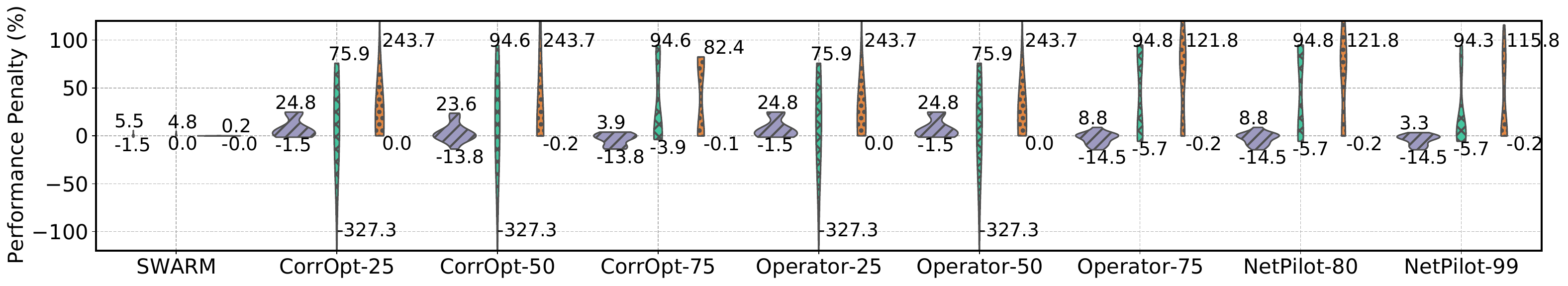}\label{fig:scenario_1_results:linear-comparator}} \\
  \subfigure[Scenario 2]{\label{scenario_2_results:linear-comparator}\includegraphics[width=0.45\textwidth]{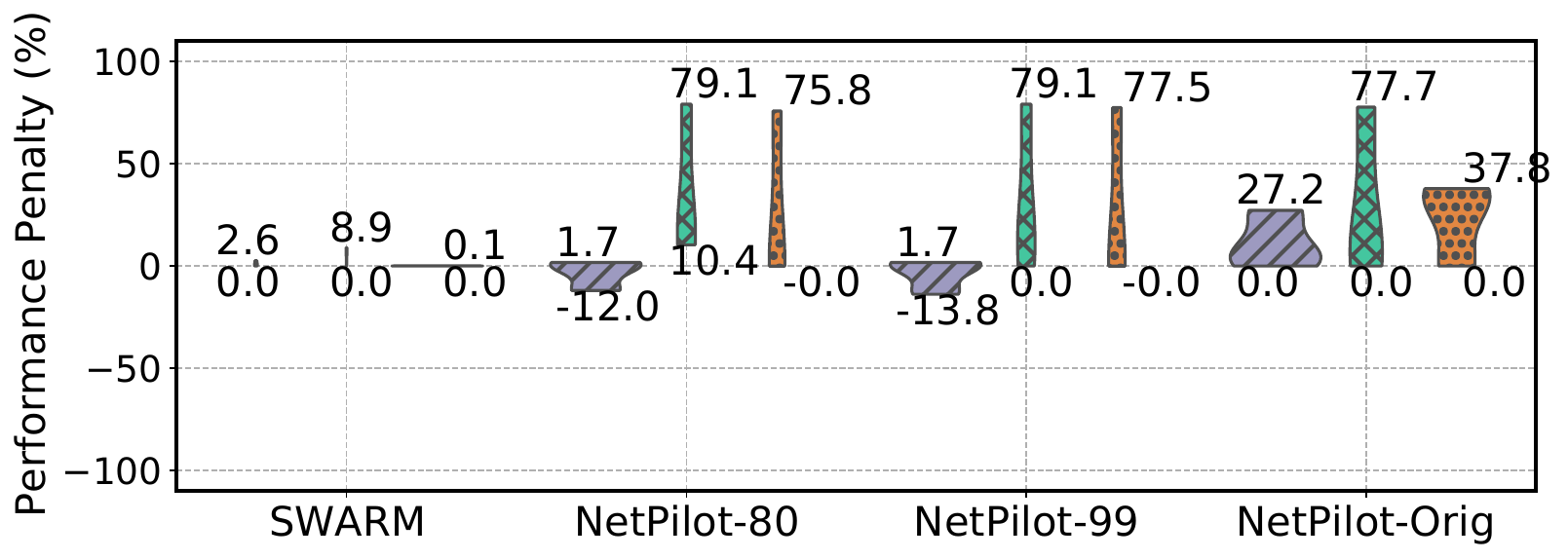}}
  \subfigure[Scenario 3]{\label{scenario_3_results:linear-comparator}\includegraphics[width=0.45\textwidth]{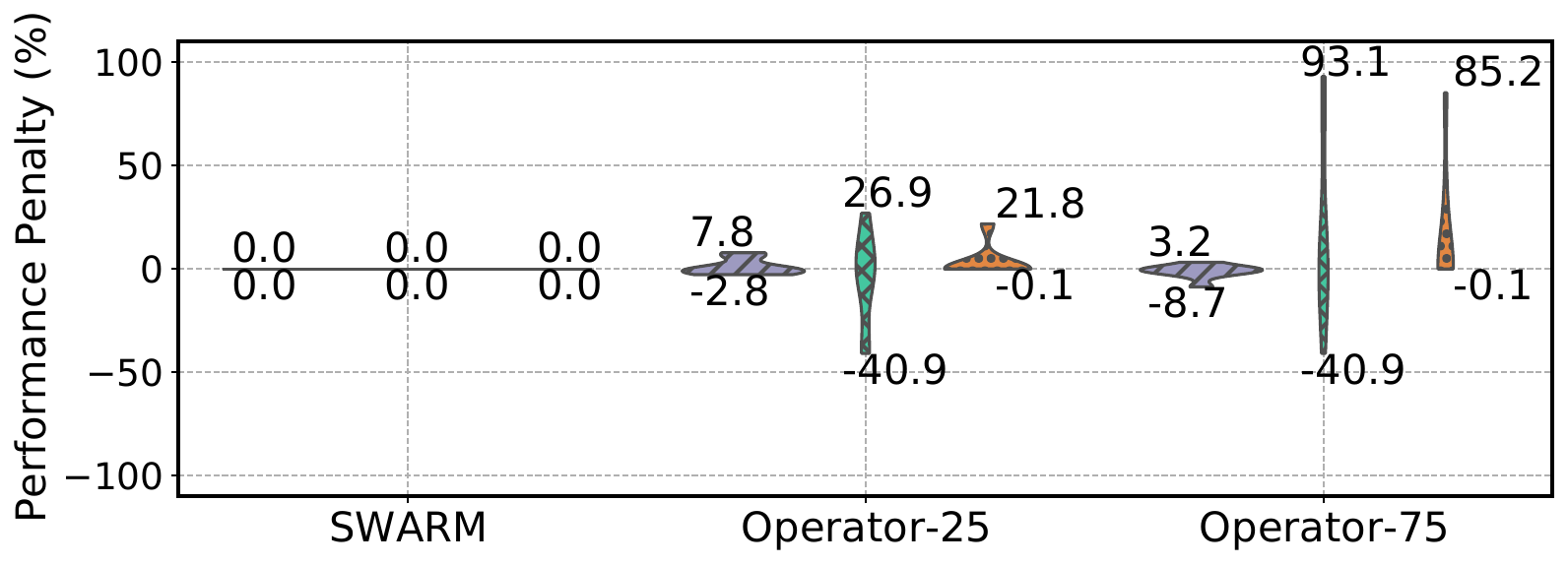}}
  \caption{\textbf{\sysname outperforms other baselines under Linear combination comparator \cadd{in Mininet}}.
  \sysname consistently achieves low penalty (always $\leq$ 8.9\%) across all the metrics and all the scenarios.}
  \label{fig:scen-linear-comparator}
\end{figure*}

\parab{Linear combination.} This comparator minimizes a weighted combination of the three CLP metrics (99p FCT, 1p throughput, and average throughput);
$$ w_0\frac{\text{99p~FCT}}{\text{99p~FCT}_{\text{h}}} + w_1\frac{\text{1p~Thru}_{\text{h}}}{\text{1p~Thru}} + w_2\frac{\text{avg~Thru}_{\text{h}}}{\text{avg~Thru}}$$ 
where $w_i$ is the assigned weight and Metric$_\text{h}$ is the metric measured in a healthy network. Note that we prefer lower FCT but higher throughput. Therefore, we use their inverses in our definition of the linear comparator. \sysname admits any combination of weights, but we evaluate for the case where all weights are set to 1. This is different from all the other comparators as it does not set any preference for any metric.

\parab{Summary of the results.} \figref{fig:scen-priority-1pT} and \figref{fig:scen-linear-comparator} compare the performance penalty of \sysname against other baselines across the three types of failures for Priority1pT and Linear combination comparator. In summary, \sysname's performance penalty is low across all the scenarios and metrics. 

\section{Disabling the Congested Device}
\label{a:logic-mitigation}

The routing protocols such as ECMP ignore the asymmetry in the datacenters, which can cause congestion. We can mitigate this by disabling the congested link or device so the routing can utilize other paths. For instance, each logical link between two switches~\cite{Wu_NetPilot} consists of multiple physical links. A control protocol multiplexes the packets over these physical links by hashing the packet's header. When a cut happens in any of these physical links, the link capacity decreases and can cause congestion on the remaining physical links. To mitigate this, we can disable the logical link to enable ECMP/WCMP to use other paths with healthy links.

\section{\sysname's Benefits}
\label{a:benefit-two-ex}

The reason behind \sysname's benefit depends on the scenario. In some incidents, \sysname chooses the actions that are already supported but ignored by prior work. In some other cases, the benefit is due to the larger action space of \sysname. We show this using two example failures on the topology in~\figref{fig:example_two_drops_motiv}.

\parab{Scenario 1 (same action space).} \cradd{In this case, a link between a tier-0 and tier-1 switch start dropping packets (\eg C0~--~B1). Existing threshold-based methods~\cite{Wu_NetPilot,Zhuo_CorrOpt} will take no action if they use a large threshold. This decision process completely ignores the impact of the packet drop rate. If the packet drop rate is high, we should disable the link. If these threshold-based methods use a smaller threshold, they end up always disabling the link. This decision completely ignores the impact of traffic load and the packet drop rate. Therefore, it can cause severe congestion. \sysname takes all these factors into account and finds much more effective mitigations (even though the action space is the same).}

\parab{Scenario 2 (larger action space).} Imagine a sequence of failures. First, a link between a tier-0 and tier-1 switch starts dropping packets at a moderate rate. \sysname would decide to disable the link. Later, another link between the same tier-0 switch and another tier-1 switch starts dropping packets at a much higher rate. In this case, \sysname disables the new link and re-enables the old link (undoing its previous action). This is because disabling both links would reduce the capacity of the network substantially, causing severe congestion. \sysname can reason about these cases, which enables exploring a broader set of actions, such as bringing back less faulty links.

\newpage
